\documentclass[usenatbib]{mn2e}
\usepackage{epsfig}
\usepackage{amsmath,amssymb}
\usepackage{graphicx}
\usepackage{epsf}

\newcommand{\hmpc}{h^{-1}\mathrm{Mpc}}
\newcommand{\hkpc}{h^{-1}\mathrm{kpc}}
\newcommand{\hMsun}{h^{-1}M_{\odot}}

\def\Mvir{M_{\rm vir}}

\def\se{\sigma_8}

\def\Ol{\Omega_{\Lambda}}

\def\Om{\Omega_m}
\def\Ob{\Omega_b}

\def\h1{h^{-1}}

\def\LCDM{$\Lambda$CDM}

\def\hMsun{h^{-1}{\ }{\rm M_{\odot}}}
\def\hMpc{h^{-1}{\ }{\rm Mpc}}

\def\hkpc{h^{-1}{\ }{\rm kpc}}

\newcommand{\mnras}{MNRAS}

\newcommand\apj{ApJ}
\newcommand\apjs{ApJS}
\newcommand\apjl{ApJL}

\title[The Shape of Dark Matter Halos]{The Shape of Dark Matter Halos: Dependence on Mass, Redshift, Radius, and Formation}

\author[Allgood et al.]
{\parbox[t]\textwidth{Brandon Allgood$^1$,
Ricardo A.~Flores$^2$,
Joel R.~Primack$^1$,
Andrey V.~Kravtsov$^3$,
Risa H.~Wechsler$^{3,4}$,
Andreas Faltenbacher$^5$, and
James S.~Bullock$^6$}
\vspace*{6pt} \\
$^1$Physics Department, University of California, 
Santa Cruz, CA 95064; {\tt allgood@physics.ucsc.edu, joel@scipp.ucsc.edu}
\\
$^2$Department of Physics and Astronomy, University of Missouri -- St.~Louis,
St.~Louis, MO 63121; {\tt ricardo.flores@umsl.edu}
\\
$^3$Dept. of Astronomy and Astrophysics, Kavli Institute for Cosmological Physics,
and The Enrico Fermi Institute,\\ $\ $The University of Chicago, Chicago, IL 60637;
{\tt andrey@oddjob.uchicago.edu, risa@cfcp.uchicago.edu}
\\
$^4$Hubble Fellow, Enrico Fermi Fellow
\\
$^5$Lick Observatory, University of California, Santa Cruz, CA 95064;
{\tt fal@ucolick.org}
\\
$^6$Center for Cosmology, Department of Physics and Astronomy, University of California,
Irvine, CA 92697; {\tt bullock@uci.edu}
\\
}
\date{\today}
\begin{document}
\maketitle

\begin{abstract} 
Using six high resolution dissipationless simulations with a varying box size
in a flat LCDM universe, we study the mass and redshift dependence of dark
matter halo shapes for $M_{\rm vir} = 9.0 \times 10^{11} - 2.0 \times 10^{14}$,
over the redshift range $z = 0 - 3$, and for two values of $\sigma_8=0.75$ and
$0.9$.  Remarkably, we find that the redshift, mass, and $\sigma_8$ dependence
of the mean smallest-to-largest axis ratio of halos is well described by the
simple power-law relation $\langle s \rangle = (0.54 \pm 0.02)(M_{\rm
vir}/M_{*})^{-0.050 \pm 0.003}$, where $s$ is measured at $0.3R_{\rm vir}$ and
the $z$ and $\sigma_8$ dependences are governed by the characteristic nonlinear
mass, $M_{*}=M_{*}(z,\sigma_8)$.  We find that the scatter about the mean $s$
is well described by a Gaussian with $\sigma \sim 0.1$, for all masses and
redshifts.  We compare our results to a variety of previous works on halo
shapes and find that reported differences between studies are primarily
explained by differences in their methodologies.  We address the evolutionary
aspects of individual halo shapes by following the shapes of the halos through
$\sim 100$ snapshots in time. We determine the formation scalefactor $a_c$ as
defined by \citet{wechsler_etal02} and find that it can be related to the halo
shape at $z = 0$ and its evolution over time.
\end{abstract}


\section{Introduction}
\label{sec:intro}

A generic prediction of cold dark matter (CDM) theory is the process of bottom
up halo formation, where large halos form from the mergers of smaller halos,
which are in turn formed from even smaller halos.  This is a violent process
and it violates most of the assumptions that go into the spherical top-hat
collapse model of halo formation which is often used to describe halos.  Since
mass accretion onto halos is often directional and tends to be clumpy, one would
not expect halos to be spherical if the relaxation time of the halos were
longer than the time between mergers and/or if the in-falling halos came along
a preferential direction (such as along a filament).  In both theoretical
modelling of CDM and observations, halos are found to be very non-spherical.  In
fact, spherical halos are rare.  Therefore, the analysis of halo shapes can
give us another clue to the nature of the dark matter and the process of halo
and galaxy formation.

One way of quantifying the shape of a halo is to go one step beyond the
spherical approximation and approximate halos by ellipsoids. Ellipsoids are
characterised by three axes, $a,b,c$, with $a \geq b \geq c$, which are
normally described in terms of ratios, $s \equiv c/a$, $q \equiv b/a$, and $p
\equiv c/b$.  Ellipsoids can also be described in terms of three classes, which
have corresponding ratio ranges: prolate (sausage shaped) ellipsoids have $a >
b \approx c$ leading to axial ratios of $s \approx q < p$, oblate (pancake
shaped) ellipsoids have $a \approx b > c$ leading to axial ratios of $s \approx
p < q$, and triaxial ellipsoids are in between prolate and oblate with $a > b >
c$.  Additionally, when talking about purely oblate ellipsoids, $a = b$, it is
common to use just $q$, since $s$ and $q$ are degenerate.

There have been many theoretical papers published over the years which examined
the subject of halo shapes. The early work on the subject includes
\citet{barnes_efstathiou87,dubinski_carlberg91,katz91,warren_etal92,
dubinski94,jing_etal95,tormen97,thomas_etal98}.  All of these works agreed that
halos are ellipsoidal, but otherwise differ in several details.
\citet{dubinski_carlberg91} found that halos have axial ratios of $s \sim 0.5$
in the interior and become more spherical at larger radii, while
\citet{frenk_etal88} found that halos are slightly more spherical in the
centres.  \citet{thomas_etal98} claimed that larger mass halos have a slight
tendency to be more spherical, where more recent simulation results find the
opposite.  Despite these disagreements many of the early authors give us clues
into the nature of halos shapes.  \citet{warren_etal92} and \citet{tormen97}
showed that the angular momentum axis of a halo is well correlated with the
smallest axis, $c$, although, as most of these authors pointed out, halos are
not rotationally supported.  This therefore has led many to conclude that the
shapes are supported by anisotropic velocity dispersion.  \citet{tormen97} took
it one step further and found that the velocity anisotropy was in turn
correlated with the infall anisotropy of merging satellites.  Most authors
found that the axial ratios of halos are $\sim 0.5 \pm 0.1$ and that halos tend
to be prolate as opposed to oblate in shape.  The most likely source of the
disagreement in these works and in the more recent works we describe below is
the different methods used often coupled with inadequate resolution.

More recent studies of halos' shape were performed by
\citet{bullock02,jing_suto02}; \citet*{springel04};
\citet{bailin_steinmetz04,kasun_evrard04,hopkins_etal05}. The results of these
authors differ, in some cases, considerably. One of the goals of this paper is
to carefully examine the differences in the findings presented by the above
authors.

All of the aforementioned publications \citep[except][]{springel04}
analyse simulations with either no baryons or with
adiabatic hydrodynamics.  This is both due to the cost associated with
performing self consistent hydrodynamical simulations of large volumes with
high mass resolution and due to the fact that very few cosmological simulations
yet produce realistic galaxies.  Nonetheless, we know that the presence of
baryons should have an effect on the shapes of halos due to their collisional
behaviour.  Three recent papers have attempted to examine the effect of baryons
\citep{springel04,kazantzidis_etal04b,bailin_etal05}.  In \citet{springel04}
the same simulations were done using no baryons, adiabatic
baryons and baryons with cooling and star formation.  In the first two cases
there was very little difference, but with the presence of cooling and star
formation the halos became more spherical.  The radial dependence of shape also
changes such that the halo is more spherical in the centre.  At $R > 0.3R_{\rm
vir}$ the axial ratio $s$ increases by $\leq 0.09$, but in the interior the
increase $\Delta s$ is as much as $0.2$.  In an independent study,
\citet{kazantzidis_etal04b} found an even larger effect due to baryonic cooling
in a set of 11 high resolution clusters.  At $R = 0.3 R_{\rm vir}$ the authors
found that $s$ can increase by $0.2-0.3$ in the presence of gas cooling.  The
extent of the over-cooling problem plaguing these simulations is still
uncertain.  This amount of change in the shape should be viewed as an upper
limit.  The most recent work on the subject is \citet{bailin_etal05}, who
concentrate more on the relative orientation of the galaxy formed at the centre
of eight high resolution halos than on the relative sphericity of the halos.
Despite this, from Figure 1 of \citet{bailin_etal05} it seems that they would
also predict an increase of $\sim 0.2$ for $s$.  It is still useful to study
shapes of halos without baryonic cooling.  Cooling and star formation in
simulations is still a very open question, making the effect of the cooling
uncertain.  We show in Paper II that our simulations without cooling match
shapes of X-ray clusters.  

Measurements of the shapes of both cluster and galaxy mass halos through varied
observational techniques are increasingly becoming available.  There have been
many studies of the X-ray morphologies of clusters
\citep*[see][]{mcmillan_etal98,mohr_etal95,kolokotronis_etal01} which can be
directly related to the shape of the inner part of the cluster halo
\citep{lee_suto03,buote_xu97}.  For a review of X-ray cluster shapes and the
latest results, see \citet{flores_etal05}, hereafter referred to as Paper II.

There has also been important new information on the shape of galaxy mass
halos, in particular our own Milky Way halo.  \citet{olling_merrifield00}
concluded that the host halo around the Galactic disk is oblate with a
short-to-long axial ratio of $0.7 < q < 0.9$.  Investigations of Sagittarius'
tidal streams have led to the conclusion that the Milky Way halo is oblate and
nearly spherical with $q \gtrsim 0.8$
\citep{ibata_etal01,majewski_etal03,martinez_delgado_etal04}.  However, by
inspecting M giants within the leading stream \citet{helmi04b} and
\citet*{law_etal05} found a best fit prolate halo with $s = 0.6$.
\citet{merrifield04} summarises the currently reliable observations for galaxy
host halo shapes using multiple techniques and find that the observations vary
a lot.

Another method for studying shapes of halos at higher redshift is galaxy-galaxy
weak lensing studies.  Analysing data taken with the Canada-France-Hawaii
telescope, \citet*{hoekstra_etal04} find a signal at a $99.5\%$ confidence level for halo
asphericity.  They detect an average projected ellipticity of $\langle \epsilon
\rangle \equiv \langle 1 - q_{2D} \rangle = 0.20^{+0.04}_{-0.05}$,
corresponding to $s = 0.66^{+0.07}_{-0.06}$, for halos with an average mass of
$8 \times 10^{11} \hMsun$.  Ongoing studies of galaxy-galaxy weak lensing
promise rapidly improving statistics from large scale surveys like the
Canada-France Legacy survey.

This paper is organised as follows: In Section \ref{sec:sims} we describe the
simulations, halo finding method, and halo property determination methods used
in this study.  In Section \ref{sec:methods} we discuss the method used to
determine the shapes of halos.  In Section \ref{sec:means} we examine the mean
axial ratios from our simulations and their dependence on mass, redshift and
$\se$.  We then examine the dispersion of the axial ratio versus mass relation.
We briefly discuss the shape of halos as a function of radius and then examine
the relationship of the angular momentum and velocity anisotropy to the halo
shape.  In Section \ref{sec:merg} we examine the relationship between the
formation history of halos and their present day shapes.  In Section
\ref{sec:comp} we compare our results to those of previous authors and explain
the sources of the differences.  In section \ref{sec:obs} we examine the
observational tests and implications of our findings.  Finally, Section
\ref{sec:conc} is devoted to summary and conclusions.

\section{Simulations}
\label{sec:sims}

\subsection{The Numerical Simulations}
\label{sec:numsim}

\begin{table*}
\begin{center}
\caption{Simulation parameters}
\small
\begin{tabular}{lllllllllll}
\hline\hline\\
\multicolumn{1}{c}{Name}&
\multicolumn{1}{c}{$\sigma_8$}&
\multicolumn{1}{c}{$\Ob$}&
\multicolumn{1}{c}{$L_{\rm box}$}  &
\multicolumn{1}{c}{$N_{\rm p}$}  &
\multicolumn{1}{c}{$m_{\rm p}$}  &
\multicolumn{1}{c}{$h_{\rm peak}$}  &
\multicolumn{4}{c}{$M_* (10^{12} h^{-1}\rm\ M_{\odot})$}
\\
\multicolumn{1}{c}{}&
\multicolumn{1}{c}{}&
\multicolumn{1}{c}{}&
\multicolumn{1}{c}{$h^{-1}\rm Mpc$} &
\multicolumn{1}{c}{} &
\multicolumn{1}{c}{$h^{-1}\rm\ M_{\odot}$} & 
\multicolumn{1}{c}{$h^{-1}\rm\ kpc$} &
\multicolumn{1}{c}{$z = 0$} &
\multicolumn{1}{c}{$z = 1$} &
\multicolumn{1}{c}{$z = 2$} &
\multicolumn{1}{c}{$z = 3$}
\\
\\
\hline
\\
L$80_{0.75}$ & $0.75$ & $0.030$ & $80$   & $512^3$ & $3.16\times 10^8$ & $1.2$
& $3.0$ & $0.11$ & $0.0046$ & $0.00027$ \\
L$80_{0.9a}$ & $0.9$  & $0.045$ & $80$   & $512^3$ & $3.16\times 10^8$ & $1.2$
& $8.0$ & $0.35$ & $0.019$ & $0.0013$ \\
L$80_{0.9b}$ & $0.9$  & $0.045$ & $80$   & $512^3$ & $3.16\times 10^8$ & $1.2$
& $8.0$ & $0.35$ & $0.019$ & $0.0013$ \\
L$200_{0.9}$ & $0.9$ & $0.030$ & $200$ & $256^3$ & $3.98\times 10^{10}$ & $5.0$
& $8.6$ & $0.41$ & $0.023$ & $0.0018$ \\
L$120_{0.9}$ & $0.9$  & $0.045$ & $120$  & $512^3$ & $1.07\times 10^9$ & $1.8$
& $8.0$ & $0.35$ & $0.019$ & $0.0013$ \\
L$120_{0.9r}$ & $0.9$ & $0.045$ & $40$ sphere  & $\sim256^3$ & $1.33\times 10^8$ & $0.9$ & $8.0$ & $0.35$ & $0.019$ & $0.0013$ \\
\\
\hline
\label{tab:sim}
\end{tabular}
\end{center}
\end{table*}

All our simulations (see Table \ref{tab:sim}) were performed with the Adaptive
Refinement Tree (ART) N-body code of \citet*{kravtsov_etal97} which implements
successive refinements in both the spacial grid and time step in high density
environments.  We analyse the shapes of halos and their merger histories in the
concordance flat {\LCDM} cosmological model: $\Om = 0.3 = 1 - \Ol$, $h = 0.7$,
where $\Om$ and $\Ol$ are the present-day matter and vacuum energy densities in
units of critical density and $h$ is the Hubble parameter in units of $100 {\rm
km\ s^{-1}\,Mpc^{-1}}$. The power spectra used to generate the initial conditions 
for the simulations were determined from a direct Boltzmann code calculation
(courtesy of Wayne Hu). 

To study the effects of the power spectrum normalisation and resolution, we
consider five simulations of the {\LCDM} cosmology.  The first simulation
(L$80_{0.75}$) followed the evolution of $512^3 = 1.34 \times 10^8$ particles
in a $80 \hMpc = 114.29 {\rm Mpc}$ box and was normalised to $\se = 0.75$,
where $\se$ is the rms fluctuation in spheres of $8 {\hMpc}$ comoving radius.
The second simulation (L$80_{0.9}$) is an exact replica of the L$80_{0.75}$
simulation with the same random number seed, but the power spectrum was
normalised to $\se = 0.9$.  The first simulation was also used to study the
halo occupation distribution and the physics of galaxy clustering by
\citet{kravtsov_etal04} and \citet{zentner_etal05a}.  Unfortunately, both of
these simulations were generated with a power spectrum which had a little more
than average power on large scales. This may happen when generating power
spectra due to cosmic variance. The simulation is still a good representation
of a volume in the Universe, but to avoid becoming non-linear on large scales,
the second simulation was stopped at $z = 0.1$. Due to the lower normalisation
of the L$80_{0.75}$ box it was allowed to run until $z = 0$.  We use these two
simulations to study the effects of the spectrum normalisation, but to achieve
better statistics and make predictions for $\se = 0.9$ at $z = 0$ we also
include another simulation of the same size and resolution (L$80_{0.9b}$).  The
fourth simulation (L$200_{0.9}$) followed the evolution of $256^3 = 1.68 \times
10^7$ particles in a $200 \hMpc = 285.7 {\rm Mpc}$ box.  The fifth simulation
L$120_{0.9}$ followed the evolution of $512^3$ particles in a $120 \hMpc =
171.43 {\rm Mpc}$ box and was normalised to $\se = 0.9$.  This simulation is
used for several purposes: firstly to achieve better statistics for rare high
mass objects, and secondly as the basis for the sixth simulation.  The sixth
simulation is a resimulated Lagrangian subregion of the L$120_{0.9}$ box
corresponding at $z = 0$ to a sphere in position space of diameter $D = 40
\hMpc$.  The initial conditions of the L$120_{0.9}$ box initially contained
$1024^3$ particles which were combined into $512^3$ particles used for the
initial simulation.  The Lagrangian subregion was then chosen and the original
higher resolution particles of mass $m_p = 1.33\times 10^8 \hMsun$ within this
region, corresponding to $1024^3$ particles in the box, were followed from the
initial time step, $z_i = 40$.  The high mass resolution region was surrounded
by layers of particles of increasing mass with a total of five particle species
in order to preserve the large scale gravitational field.  Only the regions
containing the highest resolution particles were adaptively refined. The
maximum level of refinement in the simulation corresponded to a peak formal
spatial resolution of $0.9 h^{-1} {\rm kpc}$.  For more details about the
multi-mass technique consult \citet{klypin_etal01}. The subregion was chosen
not to contain any halos above $M_{\rm vir} > 10^{13} \hMsun$ in order to
increase the statistics of isolated galaxy mass halos.

\subsection{Halo Identification and Classification}
\label{sec:halofind}

A variant of the Bound Density Maximum (BDM) algorithm is used to identify
halos and subhalos in our simulations \citep{klypin_etal99}.  The details of
the algorithm and parameters being used in the halo finder can be found in
\citet{kravtsov_etal04}.  We briefly describe the main steps in the halo finder
here.  First, all particles are assigned a density using the smooth
algorithm\footnote{To calculate the density we use the publicly available code
{\tt smooth}: {\tt http://www-hpcc.astro.washington.edu/tools/ tools.html}},
which uses a symmetric SPH (Smoothed Particle Hydrodynamics) smoothing kernel
on the 32 nearest neighbours.  Density maxima are then identified which are
separated by a minimum distance of $r_{min} = 50\hkpc$, defining the minimum
distinguishable separation of halos and subhalos.  Using the maxima as centres,
profiles in circular velocity and density are calculated in spherical bins.
Unbound particles are removed iteratively as described in
\citet{klypin_etal99}.  The halo catalogue used is complete for halos with
$\gtrsim 50$ particles.  This corresponds to a mass below which the cumulative
mass and velocity functions begin to flatten (see \citet{kravtsov_etal04} for
details).

The halo density profiles are constructed using only bound particles and they
are fit by an NFW profile \citep{navarro_etal96}:
\begin{equation} 
\rho_{NFW}(r) = \frac{\rho_s}{(r/r_s)(1+r/r_s)^2},
\label{eq:nfw}
\end{equation} 
where $r_s$ is the radius at which the log density profile has a slope of $-2$
and the density is $\rho_s/4$.  One of the parameters, $r_s$ or $\rho_s$, can
be replaced by a virial parameter ($R_{\rm vir}$,$M_{\rm vir}$, or $V_{\rm
vir}$) defined such that the mean density inside the virial radius is
$\Delta_{\rm vir}$ times the mean universal density $\rho_o(z) =
\Om(z)\rho_c(z)$ at that redshift:
\begin{equation} 
M_{\rm vir} = \frac{4\pi}{3}\Delta_{\rm vir}\rho_o R^3_{\rm vir}
\end{equation}
where $\rho_c(z)$ is the critical density, and
\begin{equation} 
\Delta_{\rm vir}(z) = \frac{18\pi^2 + 82(\Om(z)-1)-39(\Om(z)-1)^2}{\Om(z)} 
\end{equation} 
from \citet{bryan_norman} with $\Delta_{\rm vir}(0) \approx 337$ for the
{\LCDM} cosmology assumed here.  The NFW density profile fitting is performed
using a $\chi^2$ minimisation algorithm.  The profiles are binned
logarithmically from twice the resolution length (see Table \ref{tab:sim}) out
to $R_{500}$, the radius within which the average density is equal to 500 times
the critical density of the universe.  The choice of this outer radius is
motivated by \citet{tasitsiomi_etal04a} who showed that halos are well relaxed
within this radius.  The binning begins with 10 radial bins.  The number of
bins is then reduced if any bin contains fewer than $10$ particles or is
radially smaller than the resolution length.  This reduction of bins is
continued until both criteria are met.  Fits using this method have been
compared to fits determined using different merit functions, such as the
maximum deviation from the fit as described in \citet{tasitsiomi_etal04a} and
it was found that they give very similar results for individual halos.  After
fitting the halos the host halo and subhalo relationship is determined very
simply.  If a halo's centre is contained within the virial radius of a more
massive halo, that halo is considered a subhalo of the larger halo.  All halo
properties reported here are for halos which are determined to be isolated or
host halos (i.e., not subhalos).

\section{Methods of Determining Shapes}
\label{sec:methods}

There are many different methods to determine shapes of halos. All methods
model halos as ellipsoidal with the eigenvectors of some form of the inertia
tensor corresponding to the axes $c\leq b\leq a$ ($s\equiv c/a$ and $q\equiv
b/a$). The two forms of the inertia tensor used in the literature to determine
shape are the unweighted, 
\begin{equation} 
I_{ij} \equiv \sum_{n} x_{i,n} x_{j,n}
\label{eq:inertia}  
\end{equation}
and the weighted (or reduced),
\begin{equation} 
\tilde{I}_{ij} \equiv \sum_{n} \frac{x_{i,n} x_{j,n}}{r_n^2}
\label{eq:inertia2}  
\end{equation}
where \begin{equation} r_n = \sqrt{x_n^2 + y_n^2/q^2 + z_n^2/s^2},
\end{equation} is the elliptical distance in the eigenvector coordinate system
from the centre to the $n$th particle.  In both cases the eigenvalues
($\lambda_a \leq \lambda_b \leq \lambda_c$) determine the axial ratios
described at the beginning of Section \ref{sec:intro} with $(a,b,c) =
\sqrt{\lambda_a,\lambda_b,\lambda_c}$.  The orientation of the halo is
determined by the corresponding eigenvectors.

One would like to recover the shape of an isodensity surface.  The method used
here begins by determining $\tilde{I}$ with $s = 1$ and $q = 1$, including all
particles within some radius. Subsequently, new values for $s$ and $q$ are
determined and the volume of analysis is deformed along the eigenvectors in
proportion to the eigenvalues.  There are two options to choose from when
deforming the volume.  The volume within the ellipsoid can be kept constant, or
one of the eigenvectors can be kept equal to the original radius of the
spherical volume.  In our analysis of shapes, the longest axis is kept equal to
the original spherical radius.  After the deformation of the original spherical
region, $\tilde{I}$ is calculated once again, but now using the newly
determined $s$ and $q$ and only including the particles found in the new
ellipsoidal region.  The iterative process is repeated until convergence is
achieved.  Convergence is achieved when the variance in both axial ratios, $s$
and $q$, is less than a given tolerance.  

The analysis presented here begins with a sphere of $R = 0.3R_{\rm vir}$, and
keeps the largest axis fixed at this radius unless otherwise stated.  For
determining halo shapes accurately we limit our analysis to isolated halos with
$N_p \geq 7000$ within $R_{\rm vir}$.  This corresponds to $\Mvir \geq
2.21\times 10^{12} \hMsun$ for the $80\hmpc$ box simulations, $\Mvir \geq 7.49\times
10^{12} \hMsun$ for the $120\hmpc$ box simulation, and $\Mvir \geq 9.3\times 10^{11} \hMsun$
in the resimulated region of the $120\hmpc$ box.  For a discussion of our
resolution tests, see Appendix \ref{app:resolution}.

\section{Shapes As a Function of Halo Mass} 
\label{sec:means}

The simulations analysed here enable us to analyse halos spanning a mass range
from galaxy to cluster sized objects. These data provide an opportunity to
study the variation of shape and its intrinsic scatter with halo mass. Various
statistics are used to derive robust estimates of the dependence of shape on
halo-centric radius. Combining the detailed spatial and dynamical information
from the simulations we can relate quantities like angular momentum or velocity
anisotropy tensor to the shape and the orientation of the halo. In this section
we aim to present a comprehensive analysis of the properties of halos at all
redshifts. In section \ref{sec:merg} we will address evolutionary aspects of
individual halo shapes.  

\subsection{Median Relationships for Distinct Halos} 
\label{sec:median}

\begin{figure}
\centerline{\epsfxsize=3.8in \epsffile{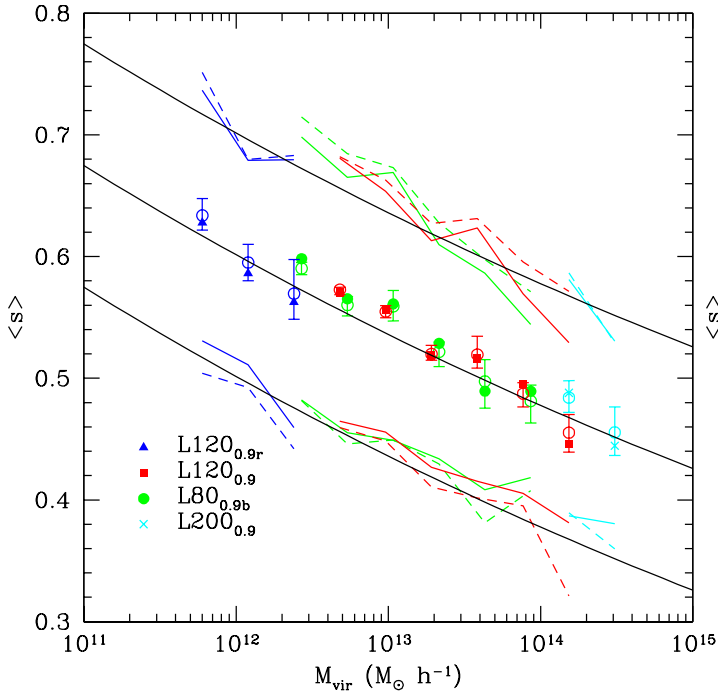}} 
\caption{Mean axial ratios $s=c/a$ for four simulations of different mass
resolution are presented with a fit (solid black lines) given by Equation
(\ref{eq:powerlaw}) and dispersion of $0.1$.  The triangles, squares, solid
circles and $\times$ symbols are the average $s$ for a given mass bin.  The
solid circles have been shifted by 0.05 in log for clarity.  The open circles
and error bars are the best fit Kolmogorov-Smirnov mean and 68\% confidence
level assuming a Gaussian parent distribution.  The dashed lines connect the
raw dispersion for each point and the coloured solid lines are the best fit (KS
test) dispersion. (See the electronic edition for colour version of the figure) 
\label{fig:means}} 
\end{figure}

We begin by fitting the mass dependence of halo shape and find that the mean
value of the axial ratio $s \equiv c/a$ decreases monotonically with increasing
halo mass as illustrated in Figure \ref{fig:means}.  In other words, less
massive halos have a more spherical mean shape than more massive halos.  Since
we use four different simulations (L$80_{0.9b}$, L$120_{0.9}$, L$120_{0.9r}$,
and L$200_{0.9}$) with varying mass and length scales we are able to determine
$\langle s \rangle(M_{\rm vir})$ over a wide mass range.  We find that over the
accessible mass range the variation of shape with halo mass is well described
by
\begin{equation} 
\langle s \rangle(M_{\rm vir},z = 0) = \alpha
\left(\frac{M_{\rm vir}}{M_{*}}\right)^{\beta} \label{eq:powerlaw}
\end{equation} 
with best fit values 
\begin{equation} \alpha = 0.54 \pm 0.03, \;
\beta = -0.050 \pm 0.003.  
\end{equation}
The parameters, $\alpha$ and $\beta$ were determined by weighted $\chi^2$
minimisation on the best fit mean data points determined via Kolmogorov-Smirnov
(KS) analysis assuming a Gaussian distribution within a given mass bin (see
section \ref{sec:meansig}).  $M_{*}(z)$ is the characteristic nonlinear mass at
$z$ such that the rms top-hat smoothed overdensity at scale $\sigma(M_{*},z)$
is $\delta_c = 1.68$.  The $M_{*}$ for $z=0$ is $8.0 \times 10^{12} \hMsun$ for
the simulations with $\Omega_b = 0.045$ and $8.6 \times 10^{12}$ for the
simulations with $\Omega_b = 0.03$. Only bins containing halos above our
previously stated lower bound resolution limit were used and only mass bins
with at least 20 halos were included in the fit.  This work extends the mass
range of the similar relationships found by previous authors
\citep{jing_suto02,bullock02,springel04,kasun_evrard04}; we compare
our results with these previous works in Section \ref{sec:comp}.

\subsection{Shapes of Halos at Higher Redshifts}
\label{sec:fixMdiffZ}

\begin{figure}
\centerline{\epsfxsize=3.8in \epsffile{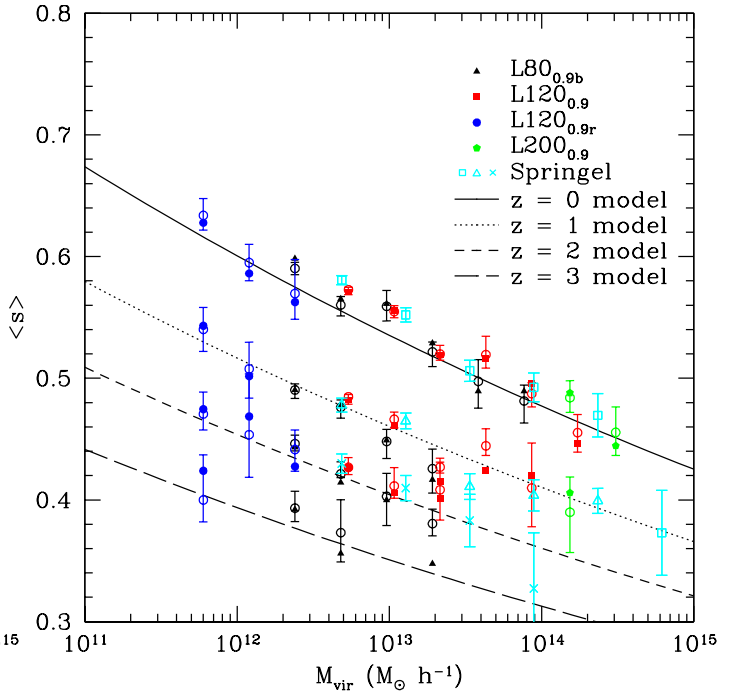}}
\caption{$\langle s \rangle(M)$ for $z = 0.0,1.0,2.0,3.0$.  The binning is
the same as in Figure \ref{fig:means}, but now for many different redshifts. 
The solid line is the power-law relation set out in Equation (\ref{eq:powerlaw}).
The L$120_{0.9}$ points are shifted by 0.05 in log for clarity.  The Springel
data agrees quite well with our data and model for $z = 0.0,1.0,2.0$.
\label{fig:evol}}
\end{figure}

The use of $M_{*}$ in the Equation (\ref{eq:powerlaw}) alludes to the evolution
of the $\langle s \rangle(M_{\rm vir})$ relation. After examining the $\langle
s \rangle(M_{\rm vir})$ relation at higher redshifts, we find that the relation
between $\langle s \rangle$ and $M_{\rm vir}$ is successfully described by
Equation (\ref{eq:powerlaw}) with the appropriate $M_{*}(z)$.  The $M_{*}$ for
$z=1.0,\ 2.0$, and $3.0$ are $3.5 \times 10^{11}$, $1.8
\times 10^{10}$, and $1.3 \times 10^9 \hMsun$ respectively for the simulations
with $\Omega_b = 0.045$.  We present our results for various redshifts in
Figure \ref{fig:evol} from the L$120_{0.9r}$, L$80_{0.9}$, L$120_{0.9}$ and
$L200_{0.9}$ simulations.  We have also included data points provided to us by
Springel (private communication) in Figure \ref{fig:evol} for comparison,
which from a more complete sample than the data presented in
\citet{springel04} and are for shapes measured at $0.4R_{\rm vir}$. 

\subsection{Dependence on $\se$}
\label{sec:sig8}

\begin{figure}
\centerline{\epsfxsize=3.8in \epsffile{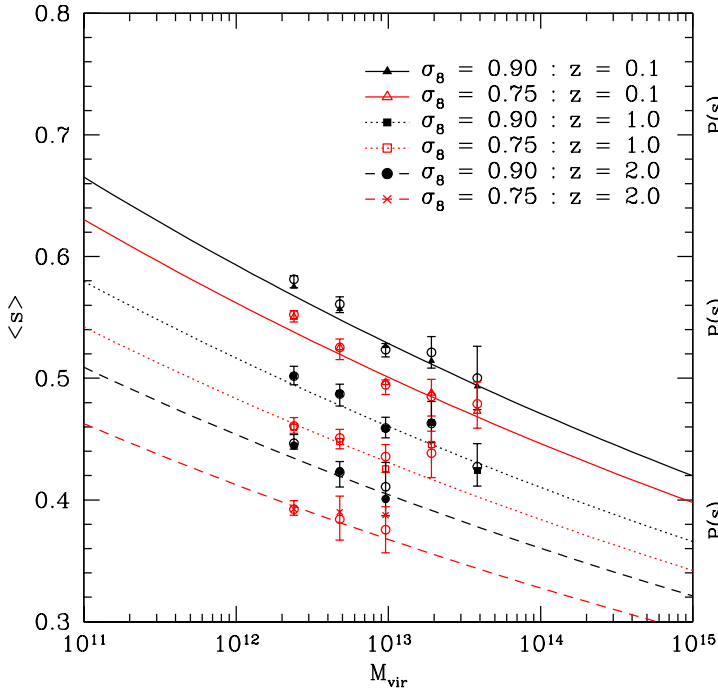}} 
\caption{$\langle s \rangle$ vs $M$ with different values of $\se$.  Different
values of $\se$ predict different values for the $\langle s \rangle$ vs $M$
relationship.  Here one can see that a universe with a lower $\se$ produces
halos which are more elongated, although the power-law relationship (Equation 
(\ref{eq:powerlaw})) remains valid, as shown by the agreement between the points 
and the lines representing this prediction.
\label{fig:sig8}}
\end{figure}

Of the parameters in the {\LCDM} cosmological model the parameter which is the
least constrained and the most uncertain is the normalisation of the
fluctuation spectrum, usually specified by $\se$. Therefore, it is of interest
to understand the dependence of the $\langle s \rangle(M_{\rm vir})$ relation
on $\se$.  Since $M_{*}$ is dependent on $\se$ the scaling with $M_{*}$ in
Equation (\ref{eq:powerlaw}) may already be sufficient to account for the $\se$
dependence. As stated in Section \ref{sec:sims}, L$80_{0.75}$ and L$80_{0.9a}$
were produced with the same Gaussian random field but different values for the
normalisation. Therefore the differences between the two simulations can only
be a result of the different values for $\se$.  As Figure \ref{fig:sig8}
illustrates, the two simulations do indeed produce different relations. We find
that the $M_{*}$ dependence in Equation (\ref{eq:powerlaw}) is sufficient to
describe the differences between simulations of different $\se$.  One should
expect this from the result of the previous subsection, that the redshift
evolution was also well described by the $M_{*}$ dependence.  The values of
$M_{*}$ for $z = 0.1$ are $5.99 \times 10^{12}$ for $\se = 0.9$ and $2.22
\times 10^{12}$ for $\se = 0.75$.  The value of $M_{*}$ for $\se = 0.75$ at
$z=1$ and $2$ are $1.09 \times 10^{11}$ and $4.57 \times 10^{9}$ respectively.
A simple fit to the redshift dependence of $M_*$ in these cosmologies is
${\rm log}(M_*) = A - B{\rm log}(1+z) - C({\rm log}(1+z))^2$, with 
$A (B, C) = 12.9 (2.68, 5.96)$ for $\sigma_8 = 0.9$ and
$A (B, C) = 12.5 (2.94, 6.28)$ for $\sigma_8 = 0.75$, and accurate
to within $1.6 \%$ and $3.1 \%$, respectively, for $z \le 3$.

\subsection{Mean - Dispersion Relationship}
\label{sec:meansig}

\begin{figure}
\centerline{\epsfxsize=3.8in \epsffile{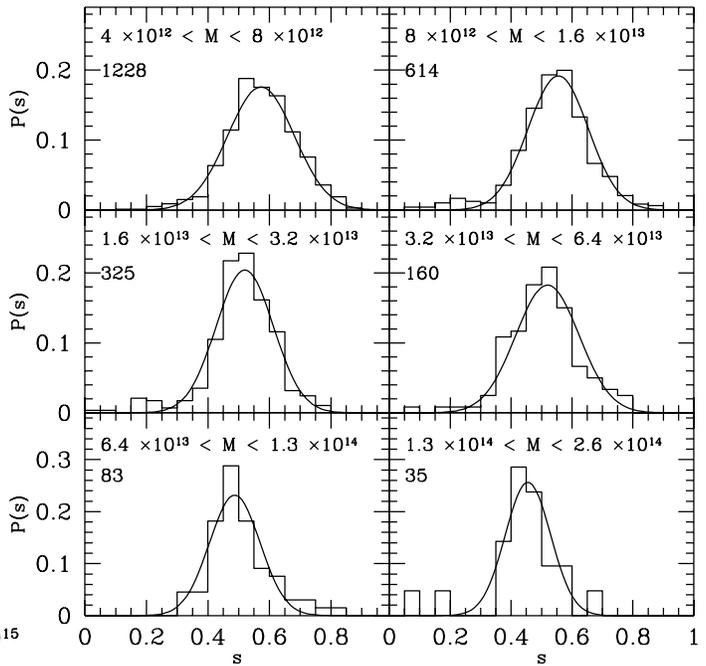}} 
\caption{The distribution of $s$ in the L$120_{0.9b}$ simulation in the mass
ranges indicated.  The number of halos in each bin is also indicated.  The
Gaussian fit shown for each graph is the best fit, based on a KS test analysis.
\label{fig:spread}}
\end{figure}

In the previous subsections we used the best KS test fit mean, assuming a
Gaussian parent distribution, as an estimate of the true mean of axial ratios
within a given mass bin.  In this subsection we examine the validity of this
assumption, and test whether the dispersion has the mass dependence suggested
by \citet{jing_suto02}.  In Figure \ref{fig:spread} we present the distribution
of $s$ in the six bins from Figure \ref{fig:means} for L$120_{0.9}$.  In each
of the plots we have also included the KS best fit Gaussian, from which the
mean was used to determine the best fit power-law in Equation
(\ref{eq:powerlaw}).  The error-bars on the mean indicated in Figure
\ref{fig:means} are the 68\% confidence limits of the KS probability.  The
limits are determined by varying the mean of the parent distribution until the
KS probability drops below 16\% for greater and less than the best fit value
for the dispersion in each mass bin. The values from this analysis
corresponding to the distributions in Figure \ref{fig:spread} can be found in
Table \ref{tab:kstest}.  In Figure \ref{fig:spread} the lowest mass bin, which
also contains the most halos, is well fit by a Gaussian.  This is seen in Table
\ref{tab:kstest} not only by the best fit KS probability, but also by the small
range of the confidence limits.  The higher mass bins are consistent with
having Gaussian parent distributions though the parent distributions' values
for the mean and dispersion are not as well constrained.  There is no
indication of a structured tail to lower values of $s$, but Table
\ref{tab:wcomp} indicates that the distributions have negative skewness.  This
arises from a small number of halos with very low values of $s$, which are
always determined to be ongoing major mergers with very close cores.

\begin{table*}
\caption{Kolmogorov-Smirnov Best Fit Values \label{tab:kstest}}
\begin{center}
\small
\begin{tabular}{llll}
\hline\hline\\
\multicolumn{1}{c}{Mass}&
\multicolumn{1}{c}{}&
\multicolumn{1}{c}{}&
\multicolumn{1}{c}{}
\\
\multicolumn{1}{c}{($\hMsun$)}& 
\multicolumn{1}{c}{$\langle s \rangle$}&
\multicolumn{1}{c}{$\sigma_s$}&
\multicolumn{1}{c}{KS Prob}
\\
\hline
\\
$4.8 \times 10^{12}$  & 0.583 (+0.003 -0.003) & 0.108 (+0.006 -0.005) & 0.80\\
$9.6 \times 10^{12}$  & 0.554 (+0.006 -0.006) & 0.110 (+0.005 -0.007) & 0.61\\
$1.92 \times 10^{13}$ & 0.518 (+0.009 -0.004) & 0.094 (+0.007 -0.013) & 0.72\\
$3.84 \times 10^{13}$ & 0.519 (+0.014 -0.013) & 0.108 (+0.016 -0.030) & 0.95\\
$7.68 \times 10^{13}$ & 0.486 (+0.005 -0.010) & 0.082 (+0.020 -0.015) & 0.78\\
$1.54 \times 10^{14}$ & 0.467 (+0.012 -0.014) & 0.073 (+0.050 -0.033) & 0.93\\
\\
\hline
\end{tabular}
\end{center}
\end{table*}

\citet{jing_suto02} found that the distribution of $s$ within a given mass bin
is Gaussian.  They found no indication of a tail or any low values of $s$.
This is most likely due to their treatment of halos with multiple cores (see
section \ref{sec:comp}).  \citet{bullock02} found a large tail to low values of
$s$ using $R = R_{\rm vir}$.  After repeating our analysis at $R = R_{\rm vir}$
we find the exact opposite.  We find even less indication of a tail than in the
distributions shown in Figure \ref{fig:spread}.  The difference is most likely
due to the centres of halos determined by the different halo finders used.
\citet{bailin_steinmetz04} find a more subtle but significant tail to lower
values of $s$.  This is most likely just a side effect of combining all mass
bins into one histogram.  If the histogram were divided into bins over smaller
ranges in mass, this tail would be seen as a consequence of the combination of
Gaussian distributions with the property that mass bins with a lower number of
halos also have lower mean values, as in Figure \ref{fig:spread}.
\citet{kasun_evrard04} find that the distribution about the mean ``is well fit
by a Gaussian'' and contains no halos with low values of $s$.  We find the same
in our spherical window analysis.  

\begin{figure}
\centerline{\epsfxsize=2.5in \epsffile{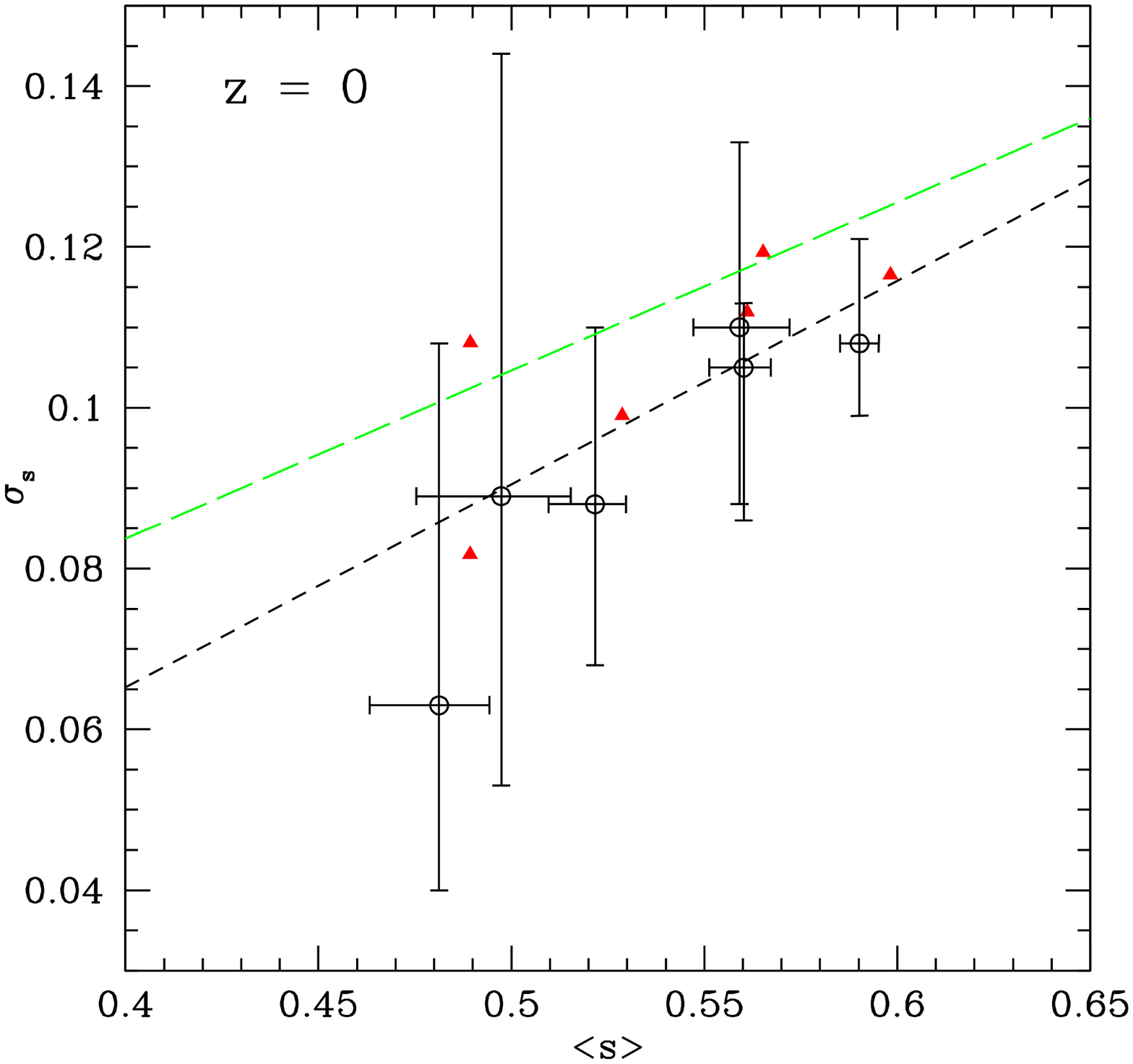}}
\centerline{\epsfxsize=2.5in \epsffile{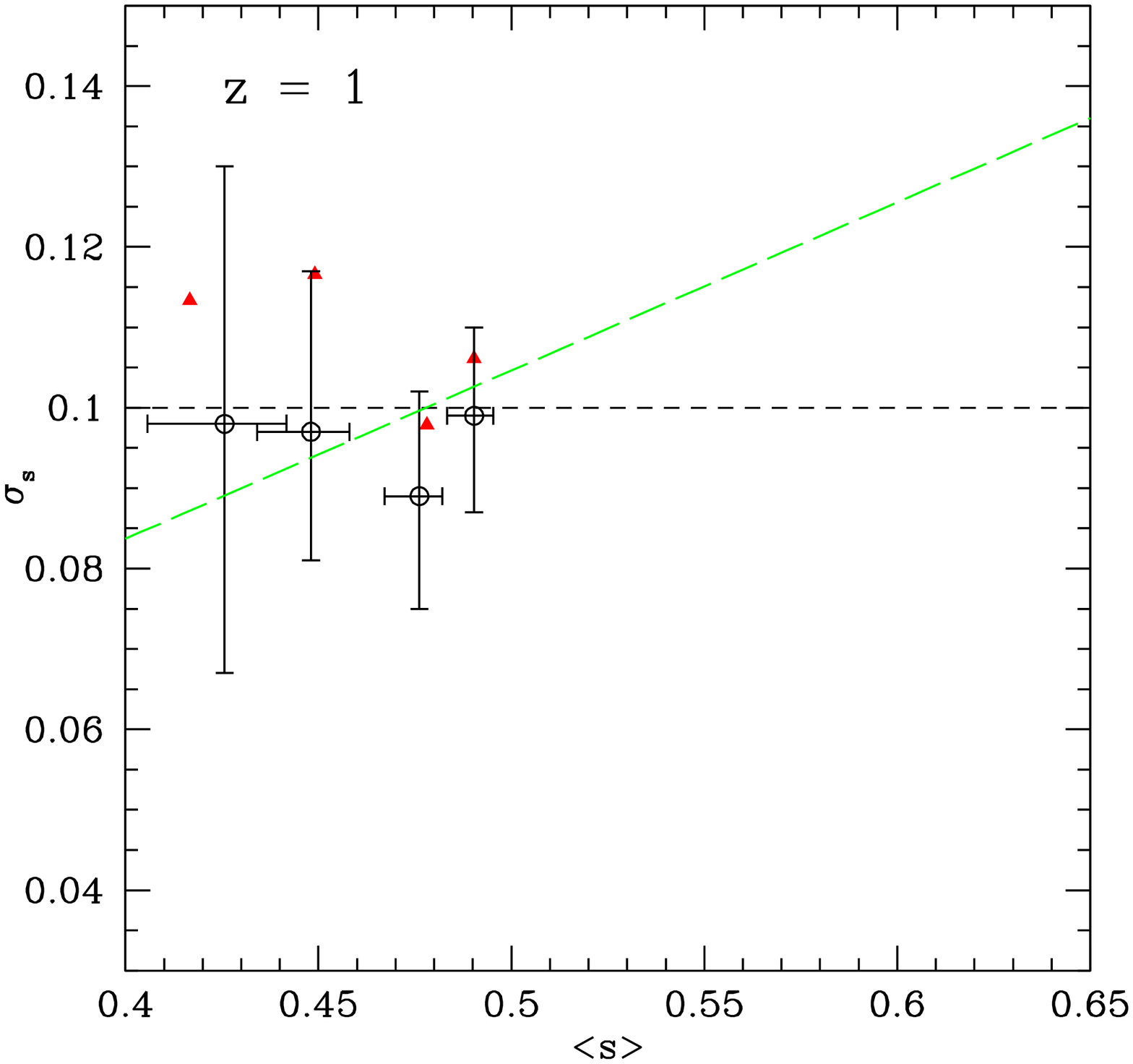}}
\caption{$\langle s \rangle$ versus $\sigma_s$.  Top: We find some evidence for
a dependence of the dispersion on $\langle s \rangle$ at $z = 0.0$ if we
perform a weighted linear least squares fit to the best values (black circles)
of $\langle s \rangle$ and $\sigma_{\langle s \rangle}$ (black short dashed
line).  It is slightly stepper than that of \citet{jing_suto02} (green long
dashed line).  Also shown are the raw average and dispersion points (red
triangles) Bottom: By $z = 1.0$ this relationship seems to have disappeared.
Due to the lack of a clear relationship between $\langle s \rangle$ and
$\sigma_s$, we favour a constant value of $\sigma_s = 0.1$, which is consistent
with all redshifts and is roughly consistent with \citet{jing_suto02} for the
values of $\langle s \rangle$ probed.
\label{fig:svssig}}
\end{figure}

Now we turn our attention to the relationship between $\langle s \rangle$ and
the dispersion for each bin.  At $z = 0$ we find a dependence of the dispersion
on $\langle s \rangle$ (top plot in Figure \ref{fig:svssig}).  The relationship
is steeper than that of \citet{jing_suto02}, who determined that $\sigma_s =
0.21 \langle s \rangle$ for the mass range they studied.  However, one can see
in the bottom plot of Figure \ref{fig:svssig} that at $z = 1$ this relationship
is no longer visible.  This may be due to the fact that the number of halos in
each bin at $z = 1$ is much lower and therefore dominated by systematics.  It
could also be that a large enough range in mass is not probed at $z = 1$ to see
the relationship.  We are unable to draw a conclusion similar to
\citet{jing_suto02}.  We therefore assume a constant value of $\sigma_s = 0.1$,
which is consistent with our results at all redshifts.

\subsection{Middle Axis Relationship}
\label{sec:p}

\begin{figure}
\centerline{\epsfxsize=3.8in \epsffile{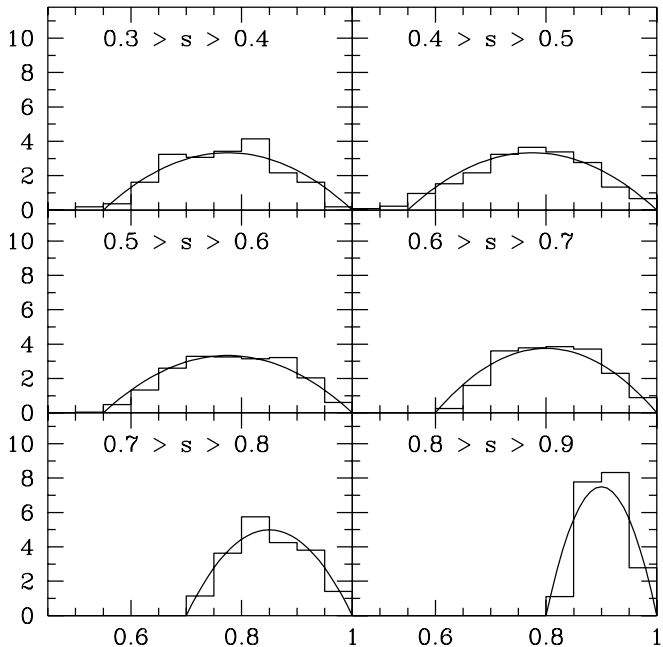}} 
\caption{The distribution of $p = c/b$ in given bins of $s$ shows a very similar
behaviour to that found in \citet{jing_suto02}.  The fit line is from Equation
(\ref{eq:jsq}), originally found in \citet{jing_suto02}.
\label{fig:qvss}}
\end{figure}

The largest to smallest axial ratio, $s$, does not uniquely determine an
ellipsoidal shape.  There is still the determination of the relationship of the
middle axis ($b$) to the smallest ($c$) or largest ($a$) axis.  We find in our
analysis, as did \citet{jing_suto02}, that the function $P(p \equiv c/b |s)$
exhibits a nice symmetric behaviour.  More commonly examined is the distribution
of $q \equiv b/a$ which can be trivially obtained from $P(p|s)$.  Figure
\ref{fig:qvss} contains six histograms of $p$ for different ranges in
$s$.  The curves are a fit proposed by
\citet{jing_suto02}, 
\begin{equation} 
P(p|s) = \frac{3}{2(1-\tilde{s})}\left[1 - \left(\frac{2p - 1 - 
\tilde{s}}{1-\tilde{s}}\right)^2\right] 
\label{eq:jsq}
\end{equation} 
with $\tilde{s} = s_{\rm min}$ for $s < s_{\rm min} = 0.55$ and
$\tilde{s} = s$ for $s \geq s_{\rm min}$.  It should also be noted that
$P(p|s) = 0$ below $\tilde{s}$.  \citet{jing_suto02} fit with a
cut-off of $s_{\rm min} = 0.5$, but otherwise we find agreement with their
results.

\subsection{Radial Dependence of Shape}
\label{sec:radius}

\begin{figure}
\centerline{\epsfxsize=3.8in \epsffile{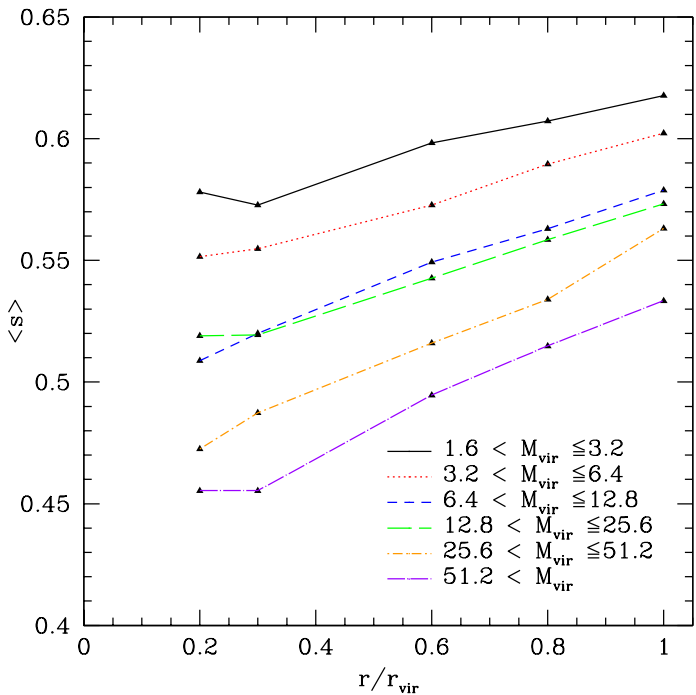}}
\caption{$\langle s \rangle$ as a function of radius at $z = 0$.
\label{fig:radshape}}
\end{figure}

The ellipsoidal shape of a halo is also found to be dependent on the radius at
which the shape is determined.  There is a systematic dependence of shape on
radius with more massive halos having a steeper gradient in $s$ with radius
than lower mass halos.  In order to study the radial dependence of shape, halos
in the $L80_{0.9b}$ simulation were examined at 5 different fractions of their
virial radius (Figure \ref{fig:radshape}).  For all halo mass bins there is a
tendency for halos to be more spherical at larger radii, with more massive
halos having a steeper change in $\langle s \rangle$ with radius.  

We also examined the value of $\langle p \rangle$ with radius and found no
radial dependence.  Therefore, $a$ and $b$ have the same radial dependence and
the largest axis $a$ becomes relatively shorter with radius.  We examine the
relationship of the radial dependence of $s$ with mass and combine it with
Equation (\ref{eq:powerlaw}) to find a shape-radius relationship,
\begin{equation}
\langle s \rangle(M_{\rm vir},r) = b(M_{\rm vir}) (r/r_{\rm vir} - 0.3) + 
0.54(M_{\rm vir}/M_{*})^{-0.05}
\end{equation}
with
\begin{equation}
b(M_{\rm vir}) = 0.037\log_{10}(M_{\rm vir}/M_{*}) + 0.062.
\end{equation}

We also examined the shape at different radii with a spherical window.  The
shape of the halo did not change very much on average with radius.  This is
consistent with the result of \citet{bailin_steinmetz04} (see Section
\ref{sec:comp}).

\subsection{Triaxiality}
\label{sec:triax}

\begin{figure}
\centerline{\epsfxsize=3.8in \epsffile{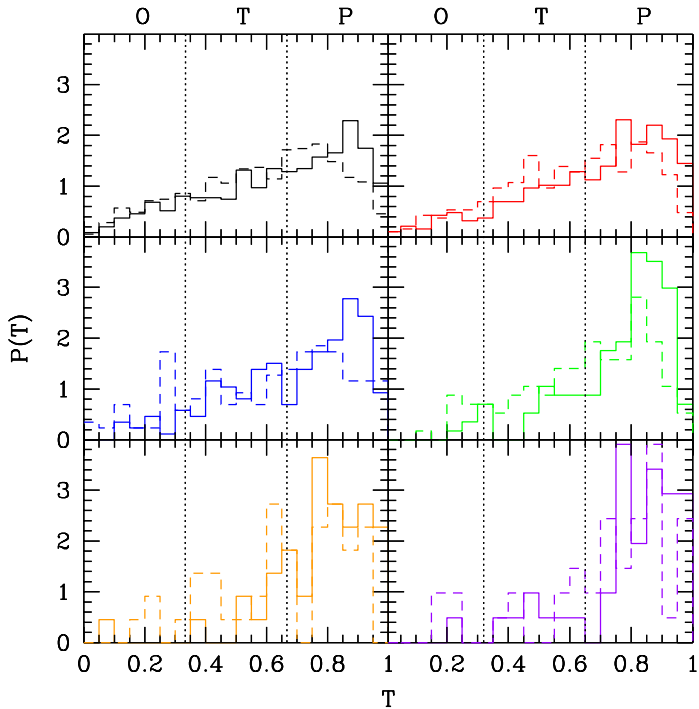}}
\caption{Triaxiality of halos at $z=0$ at $R = 0.3R_{\rm vir}$ (solid) and
$R_{\rm vir}$ (dashed).  Beginning with the top left histogram and moving
right, then down, the triaxiality of halos is divided in to the same mass bins
as in Figure
\ref{fig:radshape}.
\label{fig:triax}}
\end{figure}

Often ellipsoids are described in terms of their triaxiality (prolate, oblate,
or triaxial).  One way of expressing the triaxiality of an ellipsoid is by
using the triaxiality parameter \citet*{franx_etal91}:
\begin{equation}
T \equiv \frac{a^2 - b^2}{a^2 - c^2} = \frac{1 -q^2}{1 - s^2}.
\label{eq:triax}
\end{equation}
An ellipsoid is considered {\it oblate} if $0 < T < 1/3$, {\it triaxial} with
$1/3 < T < 2/3$, and {\it prolate} if $2/3 < T < 1$.  In Figure \ref{fig:triax}
we divide up the halos into the same mass bins as in Figure \ref{fig:radshape}
and analyse the triaxiality at $R = 0.3R_{\rm vir}$ and $R_{\rm vir}$.  We find
that most halos are prolate in shape with very few oblate halos, even at
$R_{\rm vir}$.  The deficit of halos with $T$ very close to $1$ is not
physical.  Due to the iterative process we use to define shapes, if any two of
the axes become degenerate (same length) the process has trouble converging.
In most cases it does converge but with a large systematic error. Some authors
have suggested that halos become oblate at large radii.  We find only a small
trend to less prolateness at large radii, but no evidence of a shift to oblate.
Figure \ref{fig:triax} also shows that larger halos, mainly those above
$M_{*}$, are almost entirely prolate.  Because we expect halos with masses
above $M_{*}$ to be undergoing a higher rate of merging than halos with masses
below $M_{*}$, and because it has been shown that this merging happens along
preferred directions \citep{knebe_etal04, zentner_etal05b, faltenbacher_etal05},
the prolateness is most likely due to merging.  This is in support of the idea
that halo merging is responsible for the distribution of shapes.  The internal
velocity of a halo is also related to the merger history and therefore one
would expect a relationship between he velocity structure of halos and their
shape.  We examine this in the next subsection.

\subsection{Alignment with Velocity and Angular Momentum}
\label{sec:align}

\begin{figure}
\centerline{\epsfxsize=2.5in \epsffile{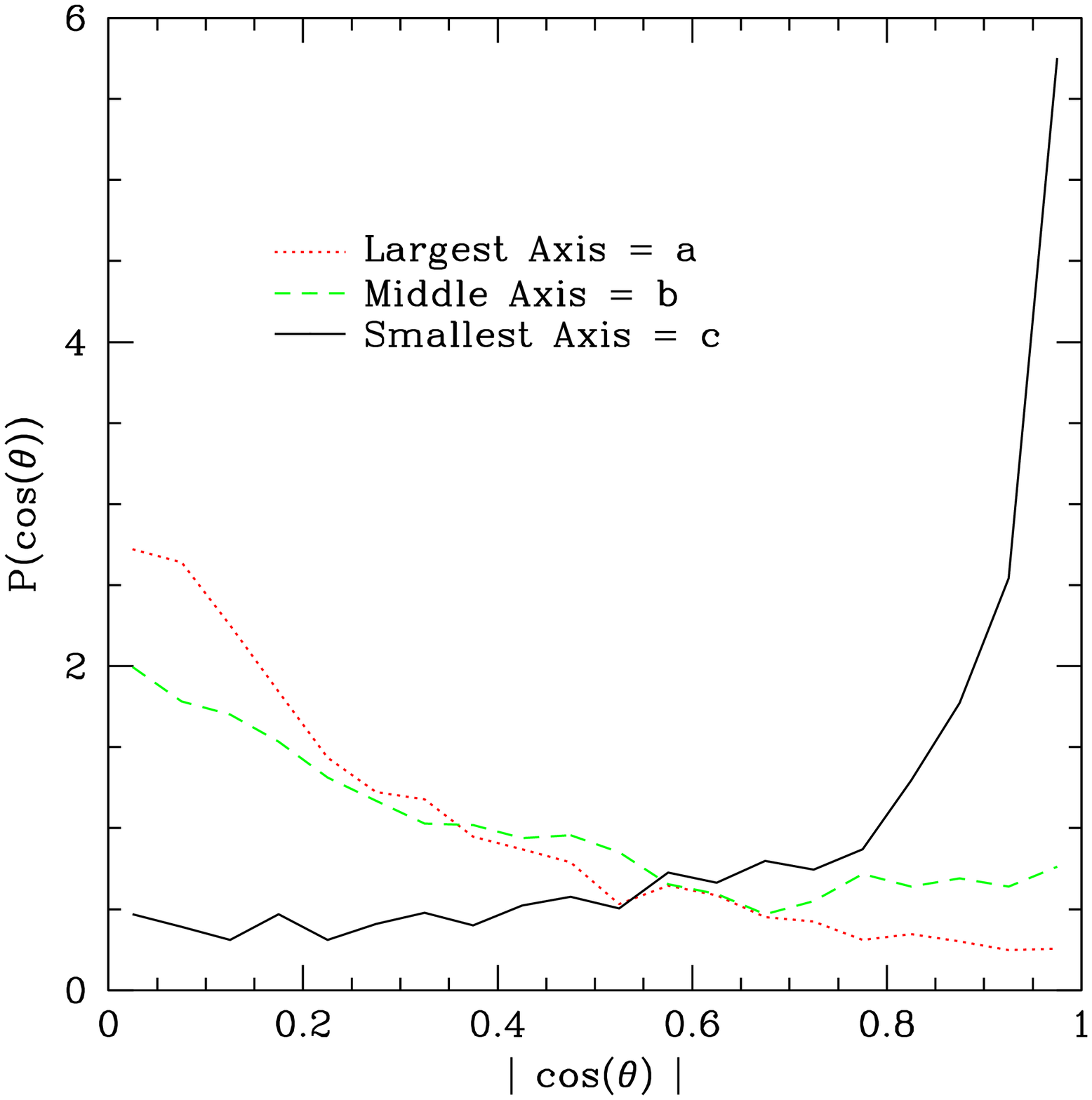}}
\centerline{\epsfxsize=2.5in \epsffile{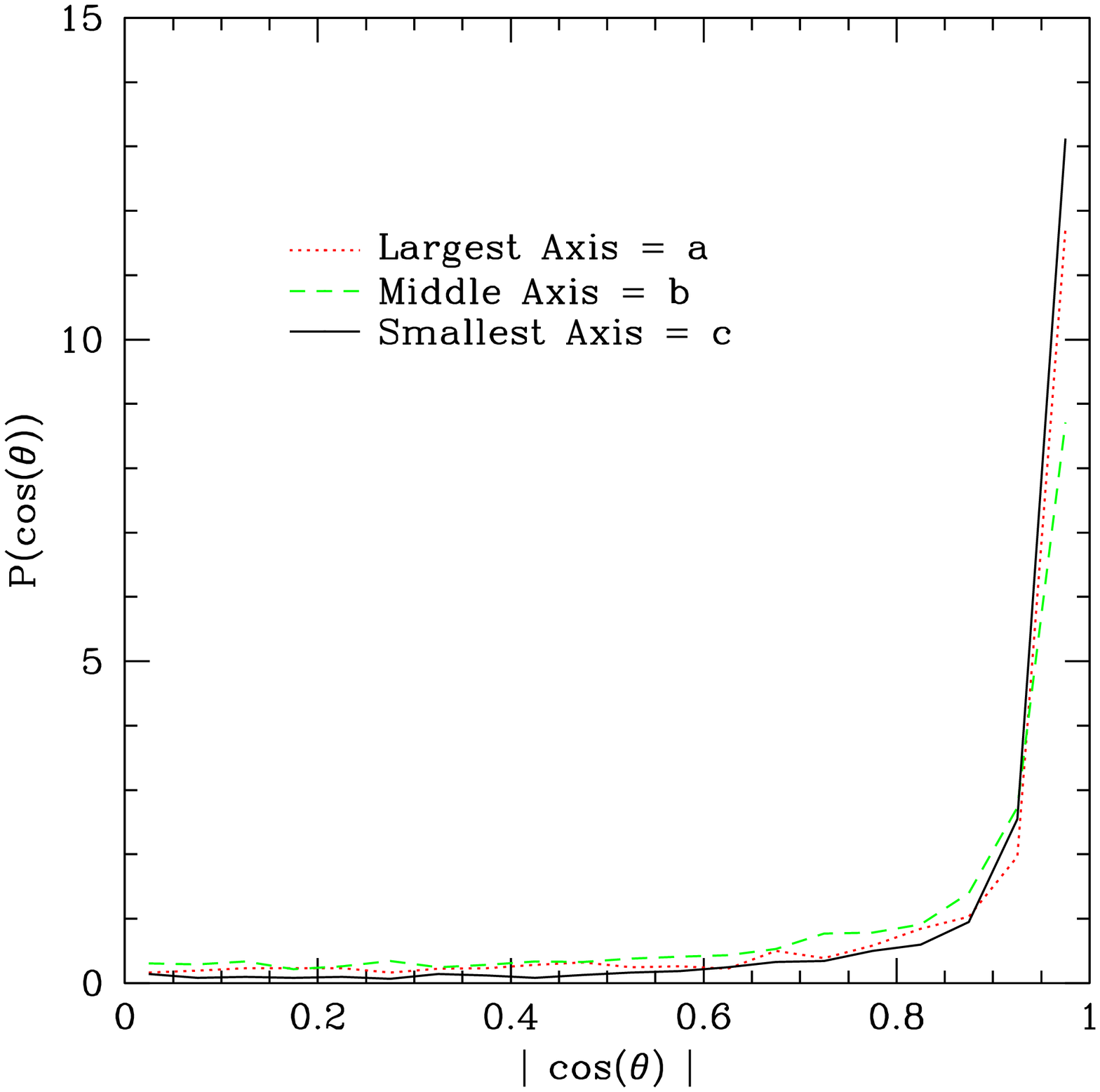}}
\caption{Top: The probability distribution of the cosine of the angles between
the largest, middle, and smallest axis and the angular momentum vector.  If the
angular momentum were randomly oriented the graph would be a flat line at a
value of $0.5$.  Bottom: The probability distribution of the cosine of the
angles between the shape axes and velocity anisotropy axes.  The velocity
anisotropy is highly correlated with the shape.
\label{fig:vel}}
\end{figure}

We have shown that the shape of a halo is related to the mass and have seen
some clues that the shape is due to merging.  Merging is also related to the
angular momentum of halos \citep{vitvitska02} and their velocity dispersion.
In order for a collisionless system such as a DM halo to sustain its shape
after merging, there must be an internal pressure provided by the velocity
dispersion.  If this is the case one would expect the internal velocities to be
correlated with the shape.  In order to investigate this we examine the
alignment of the angular momentum and the velocity anisotropy of the halo with
the shape.  The angular momentum used here is calculated using the same
particles found in the final ellipsoidal volume from our iterative method for
determining the shape, although the results do not have a large dependence on
which subset of particles within the halo is used.  We find, as was pointed out
by \citet{warren_etal92}, \citet{tormen97}, and subsequently seen by others,
that the angular momentum is highly correlated with the smallest axis of the halo.
In Figure \ref{fig:vel} we show the absolute cosine of the angle between the
indicated axis and the angular momentum vector.  If the orientations were
random the plot would be of a straight line at a value of $0.5$.  A peak at
$|\cos \theta| = 1$ means that the axes are most often aligned and a peak at
$|\cos \theta| = 0$ means that the axes are most often perpendicular to the
angular momentum.  As one can see, the smallest axis is most often aligned and the
largest axis is most often at an angle of $\pi/2$ from the angular momentum.
Although the angular momentum is aligned with the smallest axis as would be
expected for an object which is rotationally supported, DM halos are found not
to be rotationally supported.  Therefore, the significance of this alignment
points not to a cause and effect relationship but to a shared origin.  It has
been shown in previous studies that the angular momentum of halos is largely
determined by the last major merger \citep{vitvitska02}, and that, at least
during very active periods, merging (both minor and major) happens along
preferred directions \citep{knebe_etal04, zentner_etal05b, faltenbacher_etal05}.
It would seem, based on this, that the shapes and orientations of DM halos, at
least during active merging periods, can be attributed to directional merging.

In order to determine whether halos are relaxed and self supporting we examine
the relation of the velocity anisotropy to the shapes of halos. The velocity
anisotropy is defined in the same way as the inertia tensor used to measure the
shape,
\begin{equation} 
V_{ij} \equiv \sum_n v_{i,n} v_{j,n}.
\label{eq:vel}  
\end{equation}
We do not use a weighted version of this and we do not iterate, because neither
of these make much physical sense.  We calculate the velocity anisotropy tensor
again using the particles found within the ellipsoidal shell defined by the
shape analysis.  As with the angular momentum the alignment is very insensitive
to the particles used.  We then determine the angle between the respective axis
(i.e. $a,b, \&\:c$).  In Figure \ref{fig:vel} we plot the distribution of
absolute cosines between $a_{shape}$ and $a_{vel}$, $b_{shape}$ and $b_{vel}$,
and $c_{shape}$ and $c_{vel}$.  From this one can see that all three of the
axes are highly correlated.  The strength of the alignment between the velocity
anisotropy tensor and the shape suggests that the shape is supported by
internal velocities.  But how does the shape relate more directly to the
velocity anisotropy?

\begin{figure}
\centerline{\epsfxsize=3.8in \epsffile{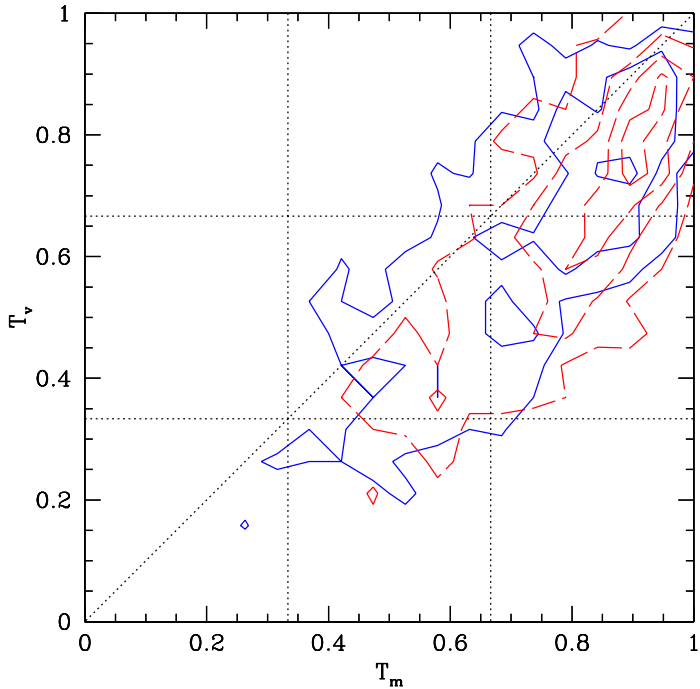}}
\caption{Triaxiality of the velocity anisotropy tensor and the shape for 
$R = 0.3R_{\rm vir}$ (blue) and $R_{\rm vir}$ (red).
\label{fig:tsupport}}
\end{figure}

In Figure \ref{fig:tsupport} we show the triaxiality of the velocity anisotropy
and the density.  The velocity anisotropy on average is more spherical in shape
than the density.  This is the expected trend from the Jeans equation for an
ellipsoidal distribution.  The velocity anisotropy is directly related to the
potential which has the same orientation as the shape but is more spherical due
to the fact that potential is related to the spacial derivative of the density.
It would therefore seem that the mass dependence of shape can not be explained
by different relaxation times.  This is also supported by the fact that halos
are more aspherical in the centre where the relaxation time would be shorter
than at the virial radius.

\section{Merger History and Shapes}
\label{sec:merg}

So far we have investigated the evolution of halo shapes in fixed mass bins as
a function of redshift and for two different values of $\se$.  Additional
insight into the origin of shapes and their dispersion can be gained by tracing
the evolution of individual halos.  In order to quantify the evolution or mass
accretion history (MAH) of the halos we have constructed merger trees.  For
more information on the merger trees, please see \cite{allgood_thesis}.  From
these merger trees we determine the MAH for each halo at $z = 0$ by following
the evolution of its most massive progenitors.  \citet{wechsler_etal02} showed
that the MAH of a halo can usually be well fit by a single parameter model, 
\begin{equation}
M(a) = M_{\rm o} \exp\left[-2a_c\left(\frac{1}{a} - 1 \right)\right],
\label{eq:ac}
\end{equation}
where $M_o$ is the mass of the halo a $z = 0$ and $a_c$ is the scale factor at
which the log slope of the MAH is $2$.  Although in the \citet{wechsler_etal02}
they only allowed $a_c$ to be a free parameter, we find that by also allowing
$M_{\rm o}$ to be a free parameter we are able to better recover $a_c$ for
halos which had experienced a recent major merger.  Halos with lower values of
$a_c$ formed earlier, and as shown by \citet{wechsler_etal02}, have a higher
concentration. By means of Equation (\ref{eq:ac}) we assign an $a_c$ to every
halo found at $z = 0$.  

\begin{figure}
\centerline{\epsfxsize=3.8in \epsffile{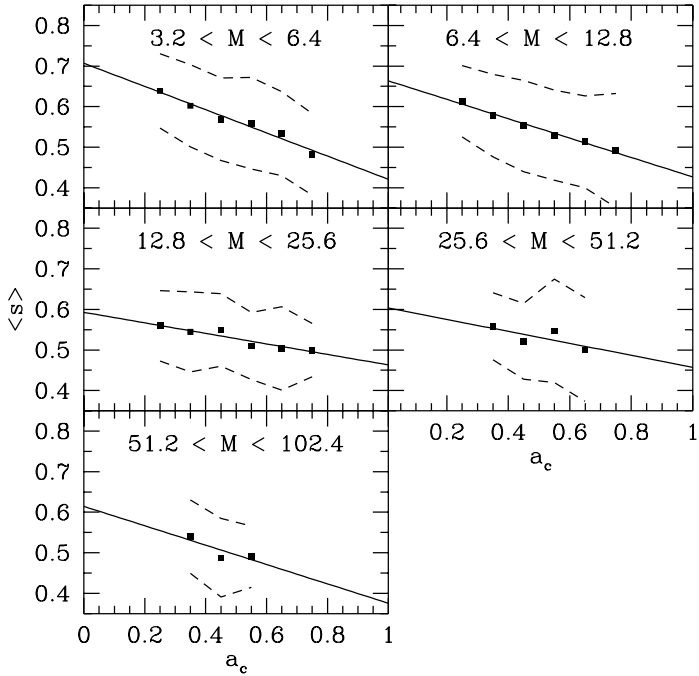}}
\caption{$\langle s \rangle$ vs characteristic formation epoch for different
mass bins (mass quoted in units of $10^{12} \hMsun$).  Only bins that contain
at least 10 halos are shown (square points).  There is a distinct trend of
shape with $a_c$ for the lower mass bins.  At higher mass there is still a
trend but it is uncertain how strong the trend is due to the lower number
statistics.  Solid black line is a linear fit to the points and dashed line is
the $1\sigma$ scatter about the points.  \label{fig:avesvsac}}
\end{figure}

\begin{figure}
\centerline{\epsfxsize=3.8in \epsffile{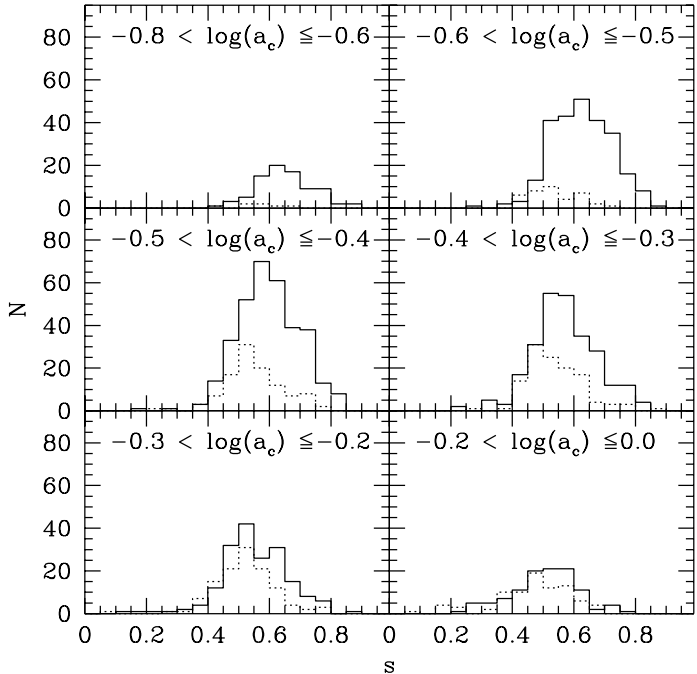}}
\caption{Distributions of $s$ for a given range in $a_c$ and mass.  The solid
histogram is for the mass range: $3.2 \times 10^{12} < M_{\rm vir} < 6.4 \times
10^{12}$ and the dotted histogram is for a mass range of $1.28 \times 10^{13} <
M_{\rm vir} < 5.12 \times 10^{13}$.  The dispersion in $s$ in a given mass bin
can be explained in part by the different MAHs.
\label{fig:acScat}}
\end{figure}

In Figure \ref{fig:avesvsac} we plot $s$ versus $a_c$ for the halos in the
L$120_{0.9}$ simulation split into separate mass bins. We find that halos which
formed earlier are on average more spherical with a dispersion of $0.08-0.1$
(see Figure \ref{fig:acScat}) for all mass bins.  This implies that the scatter
in the $\langle s \rangle(M_{\rm vir})$ relation can not be completely
attributed to the different values of $a_c$ for that particular mass bin.
However, the dependence on $a_c$ is less for the higher mass halos and this
would explain the mean-dispersion relationship explored in Subsection
\ref{sec:meansig}.  Higher mass halos were found to have a smaller dispersion
than lower mass halos at $z=0$.  This can also be seen in Figure
\ref{fig:acScat}.  It is very likely that the residual scatter is due, at least
in part, to the pattern of infall.  Since $s$ is derived from an inherently
three dimensional quantity, namely the inertial tensor, a one dimension
parameterisation may not be sufficient to capture all of the physics involved.
A more careful study of infall is needed to explain the dispersion in $s$. 

\begin{figure}
\centerline{\epsfxsize=3.8in \epsffile{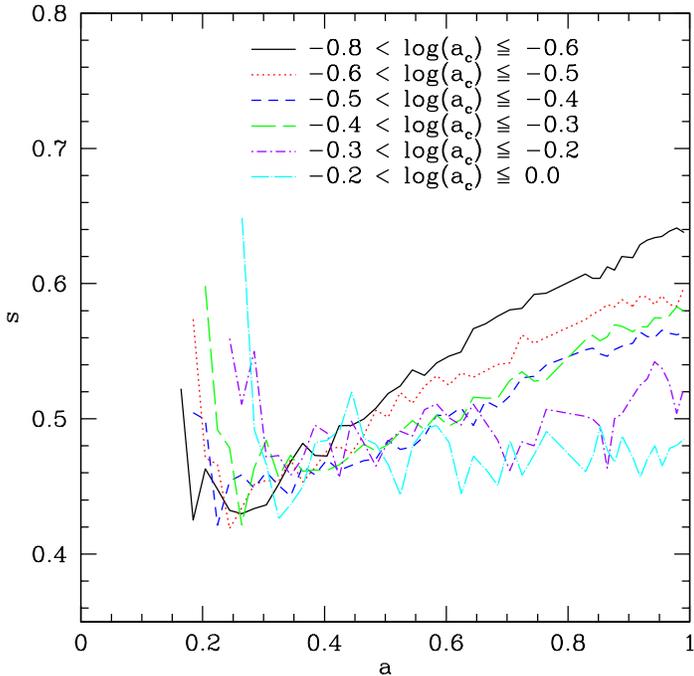}}
\caption{The evolution of halos with different values of $a_c$ in the mass bin
$3.2 \times 10^{12} < M_{\rm vir}(z=0) < 6.4 \times 10^{12}$.  Halos become
more spherical after a short period after $a_c$. The halos which form earlier
become spherical more rapidly. Log binning was chosen to even out the number of
halos in each bin.
\label{fig:acEvolve}}
\end{figure}

The above investigation makes clear that the dispersion of halo shapes cannot
be explained by appealing to a single parameter description of the MAHs.
However, an average evolutionary pattern for halos which is dependent on both
$a_c$ and mass is seen.  Figure \ref{fig:acEvolve} displays the evolution of
$\langle s \rangle$ (sorted by $a_c$) with scale factor $a$ for a particular
mass bin at $z = 0$.  Halos that formed early (lower $a_c$) are more spherical
today as was pointed out above. Moreover, they become spherical more rapidly
(indicated by the increasing slopes for halos of low $a_c$), although the
transformation rate towards spherical shapes seems to slow for all values of
$a_c$ with increasing expansion factor.  In Figure \ref{fig:acEvolve} the
results for the lowest mass bin ($3.2 \times 10^{12} < M_{\rm vir} < 6.4 \times
10^{12}$) are shown. Apart from a systematic shift to lower values of $s$ the
corresponding plots for higher mass bins look very similar.  

\begin{figure}
\centerline{\epsfxsize=3.8in \epsffile{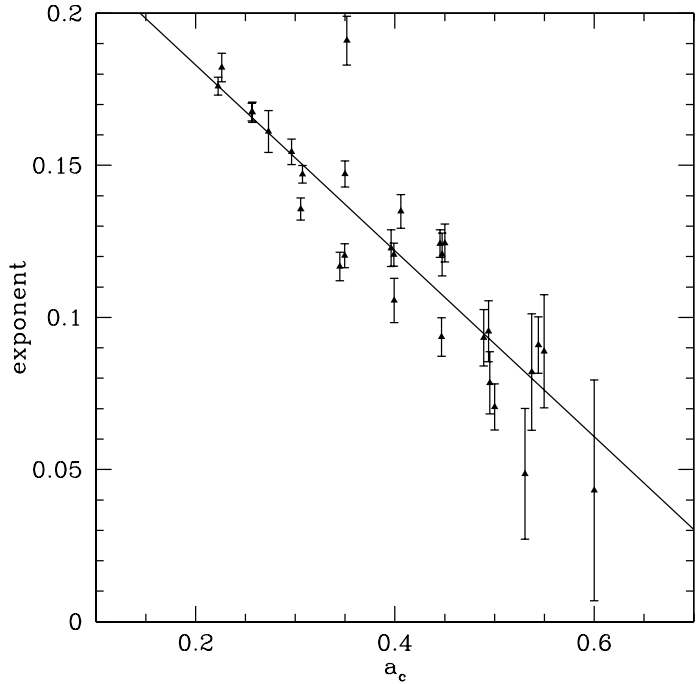}}
\caption{The rate of change exponent $\nu$ (see Equation (\ref{eq:acExp}))
versus expansion factor at halo formation $a_c$. The solid line displays the
fitting formula given by Equation (\ref{eq:nu}).  Only bins containing at lease
20 halos is displayed.  The $a_c$ value of each point is the average value for
that respective bin and the errorbars represent the variance determined by the
$\chi^2$ fitting.
\label{fig:acExp}}
\end{figure}

Halos which have early formation times (low $a_c$) at a fixed mass today have
typically accreted more mass since $a_c$ than halos with higher values of
$a_c$.  The rapid transformation towards spherical shapes for early forming
halos implies either that lower mass halos become spherical more rapidly after
$a_c$, probably due to shorter dynamical times, or that mass accretion after
$a_c$ is more spherical, therefore causing the halo to become more spherical as
well.  By examining other mass bins we find that halos of different masses
today but with comparable values for $a_c$, thus different masses at $a_c$,
show the same rate of change in $s$, but with different initial values of $s$.
This finding suggests that the rate at which a halo becomes spherical depends
on its $a_c$ rather than on its mass. We find that we can approximate the
dependence of $s$ on the expansion factor $a$ for $a > a_c + 0.1$ by a simple
power law 
\begin{equation}
\label{eq:acExp}
\langle s \rangle(a) \propto (a-a_c)^{\nu}\ :\ a > a_c + 0.1, 
\end{equation}
where $\nu$ has to be fitted for the particular halo. In Figure \ref{fig:acExp}
we display the values of $\nu$ versus $a_c$ determined by $\chi^2$ fitting for
the L$80_{0.9b}$ simulation.  The L$80_{0.9b}$ was divided into bins of log
mass and $a_c$.  The average MAH for each bin was fit by Equation
(\ref{eq:acExp}) using  $\chi^2$ minimisation.  All bins containing at least 20
halos were used determine the function $\nu$.  We find a tight correlation
between the $\nu$ and $a_c$ which can be approximated by 
\begin{equation}
\label{eq:nu}
\nu = 1.74 \times a_c^{-0.3}\ . 
\end{equation}
This fit is represented by the solid line in Figure \ref{fig:acExp}. The
remarkable success of Equation (\ref{eq:acExp}) to fit the data supports the
idea, that the transformation from aspherical to spherical halo shapes is
driven by mass accretion becoming more spherical after $a_c$.  The physical
reason for the observed behaviour merits further investigation.  

\section{Comparison with previous determinations of halo shape}
\label{sec:comp}

In the previous sections we have explored many aspects of halo shapes.  Central
to this discussion has been $\langle s \rangle(M_{\rm vir})$.  This relationship
has been examined by many recent studies all of which seem to determine different
relationships.  In this section we address these discrepancies.

\begin{figure}
\centerline{\epsfxsize=3.8in \epsffile{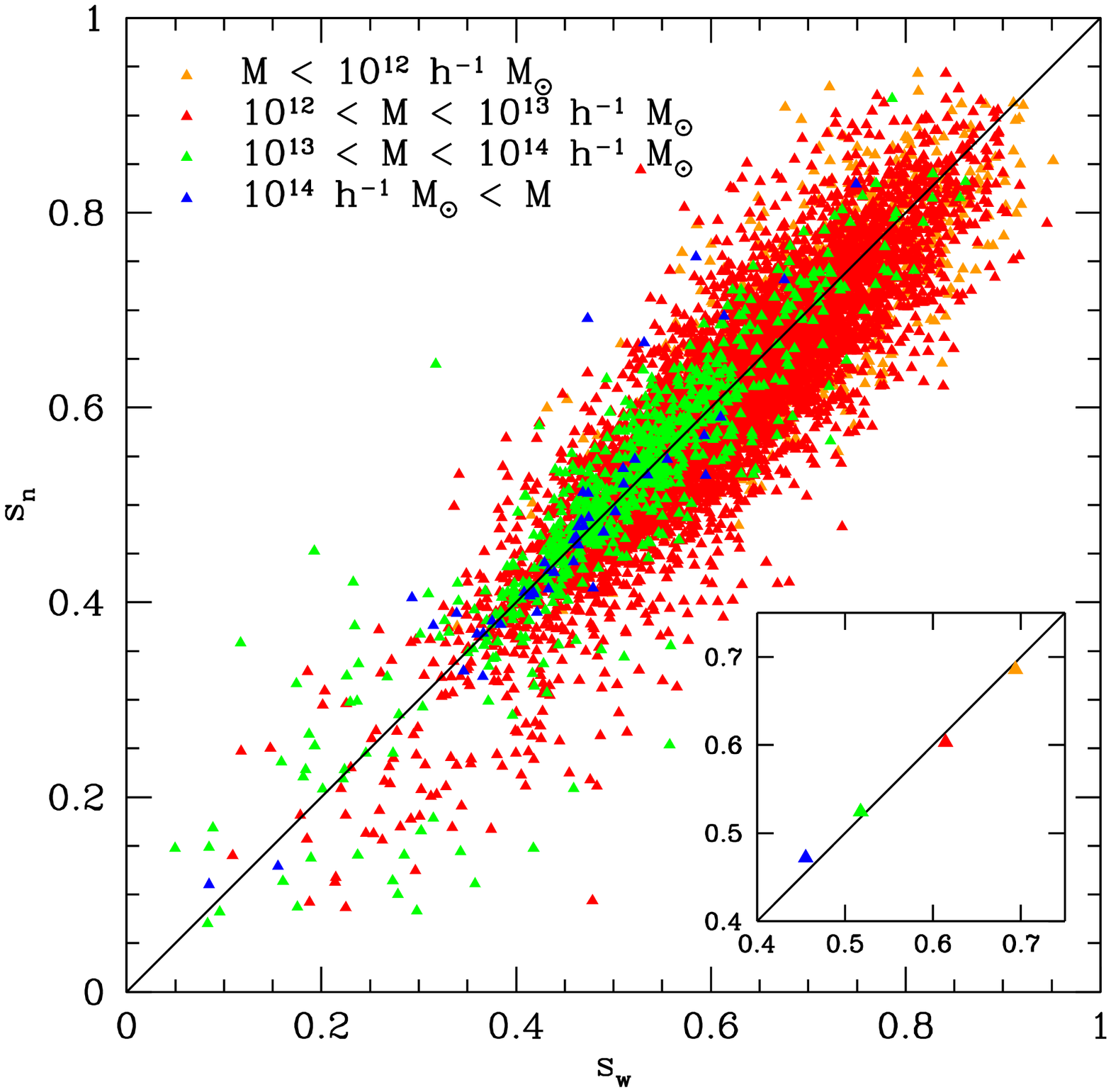}} 
\caption{Axial ratios for halos in a cosmological simulation are divided into
mass bins and the shapes are calculated using the weighted ($s_w$) and
non-weighted ($s_n$) iterative inertia tensor methods.  The two methods agree
within $\sim 10\%$ and give the same value when averaged over a given mass bin
(inset graph).
\label{fig:w_and_now}} 
\end{figure} 

First, an examination of the difference in the inferred shape from the use of
the weighted versus the unweighted inertia tensor is needed.  Most recent
authors prefer the weighted (or reduced) inertia tensor (Equation
(\ref{eq:inertia2})) which is the method we have chosen to use.  The motivation
for the use of the weighted inertia tensor, $\tilde{I}$, is due to the bias
present in the unweighted method to particles at larger radii.  By weighting
the contribution from each particle in the sum by the distance to the particle
squared, $\tilde{I}$ is less sensitive to large substructure in the outer
regions of the analysis volume.  To test the difference between the methods we
examined a sample of halos from the L$120_{0.9}$ simulation using the iterative
method with both versions of inertia tensor (Figure \ref{fig:w_and_now}).  Both
iterative methods give similar results for the mean quantities (inset in Figure
\ref{fig:w_and_now}) as a function of mass with individual halos differing by
$\Delta s \leq 0.15$.  The detailed distributions are different and have some
interesting features (Table \ref{tab:wcomp}).  We find that halos which have
lower $s$ values for the weighted method over the unweighted method have larger
substructure near the centre and halos which have lower $s$ values for the
unweighted method had larger substructure near the outer edge of the analysis
volume.  At very small axial ratios the unweighted method seems to always give
larger axial ratios. As we have shown, halos with $s \lesssim 0.3$ are late
forming and are strongly contaminated by substructure.  This leads to the
unweighted method giving a lower value of $s$ indicated by the high skewness
shown for the method in Table \ref{tab:wcomp} by the skewness.  For the two
well resolved mass ranges ($10^{12} < M_{\rm vir} < 10^{13}$ and $10^{13} <
M_{\rm vir} < 10^{14}$) the unweighted method has a larger negative skewness.
Note that all mass bins except for the last unweighted bin have negative
skewness (this was discussed in Section \ref{sec:means}), but the most massive
bin suffers from low statistics. The distribution of $\sigma_s$ values is
always broader for the unweighted method.

\begin{table}
\caption{Weighted vs. Unweighted Shapes}
\small
\begin{tabular}{llllll}
\hline\hline\\
\multicolumn{1}{c}{Method}&
\multicolumn{1}{c}{Mean Mass}&
\multicolumn{1}{c}{$\langle s \rangle$}&
\multicolumn{1}{c}{$\sigma_s$}&
\multicolumn{1}{c}{kurtosis} &
\multicolumn{1}{c}{skewness}
\\
\hline
\\
weighted & $5 \times 10^{11}$ & 0.694 & 0.094 & -0.029 & -0.132\\
unweighted & $5 \times 10^{11}$ & 0.686 & 0.097 & -0.302 & -0.053\\
weighted & $5 \times 10^{12}$ & 0.614 & 0.111 & 0.253 & -0.188\\
unweighted & $5 \times 10^{12}$ & 0.603 & 0.118 & 0.605 & -0.287\\
weighted & $5 \times 10^{13}$ & 0.518 & 0.117 & 1.756 & -0.532\\
unweighted & $5 \times 10^{13}$ & 0.524 & 0.126 & 1.628 & -0.592\\
weighted & $5 \times 10^{14}$ & 0.455 & 0.114 & 2.066 & -0.515\\
unweighted & $5 \times 10^{14}$ & 0.472 & 0.133 & 1.428 & 0.150\\
\\
\hline
\label{tab:wcomp}
\end{tabular}
\end{table}

\begin{figure}
\centerline{\epsfxsize=3.8in \epsffile{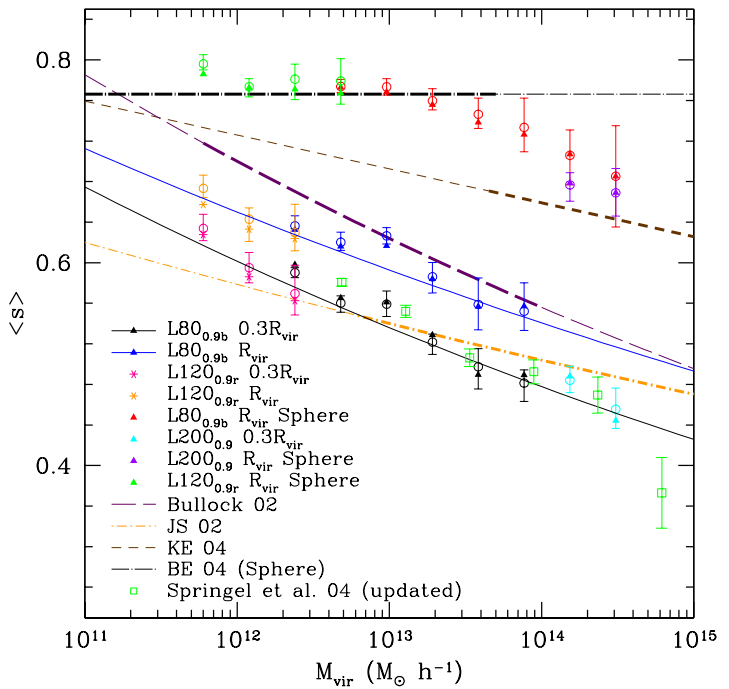}} 
\caption{Comparison of $\langle s \rangle(M_{\rm vir})$ relation with previous
studies.  We attempt to reconcile the differences between our results and those
of other authors.  We present the results of a shape analysis of the
L$80_{0.9b}$, L$120_{0.90r}$, and L$200_{0.9}$ simulations using the iterative
inertia methods at $R = 0.3R_{\rm vir}$ (black,pink,cyan) and non-iterative
spherical window analysis at $R_{\rm vir}$ (red,green,violet).  In addition, we
present the results of a shape analysis of the L$80_{0.9b}$ and L$120_{0.90r}$
using the iterative method at $R_{\rm vir}$ (blue).  The black line is our
proposed fit from Equation (\ref{eq:powerlaw}) and this should be compared to
the results of Springel (private communication) (green open squares) and
\citet{jing_suto02} (orange dot dash).  The blue line is a fit to the blue
points, which should be compared to the \citet{bullock02} line (violet long
dash).  Finally the red line is a renormalised version of the
\citet{kasun_evrard04} fit which should be compared their fit (brown small
dash). The thin black dot dash line at $\langle s \rangle \sim 0.7$  is the
spherical shell fit of \citet{bailin_steinmetz04}.  The bold portions of the
lines indicate the mass ranges where the fit was compared to simulated data by
the respective authors. \label{fig:compared}} 
\end{figure}

There are more differences than just the form of the inertia tensor used.  In
order to compare our results to a selected number of previous results (Figure
\ref{fig:compared}) we have repeated the shape analysis using the methods
described in the corresponding papers.  The previous work which is most similar
to the current work is that of \citet{springel04}.  Our findings are very
similar to the \citet{springel04} results except we found that $\langle s
\rangle(M_{\rm vir})$ has a slightly higher normalisation.  Through private
communication with Volker Springel, we were provided with an updated set of
data points which come from a more complete sample and are in much better
agreement with our results (open green squares in Figure \ref{fig:compared}).
Not only do our results agree at $z = 0$, but also at higher redshift (see
Figure \ref{fig:evol}).

\begin{figure}
\centerline{\epsfxsize=3.8in \epsffile{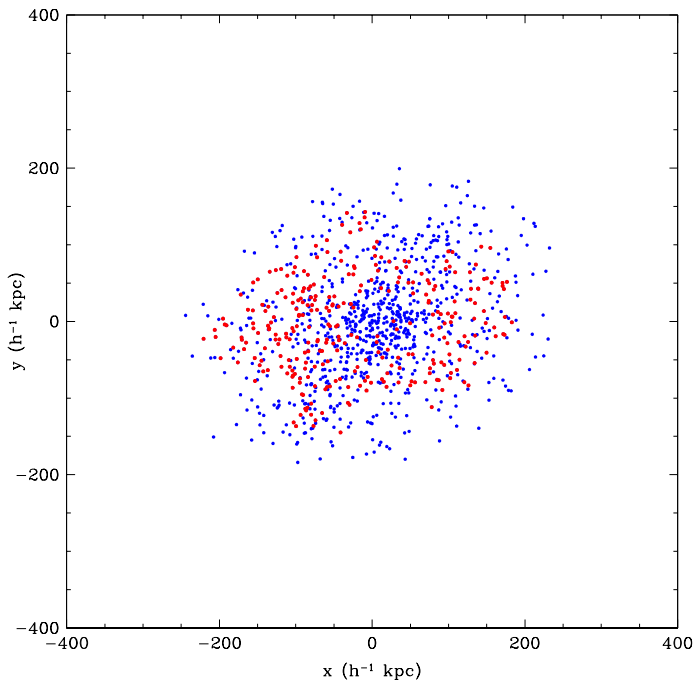}} 
\caption{Comparison of an isodensity shell (red) and a tensor ellipsoid
(blue). Particles are selected by the JS (red) and inertia tensor (blue)
methods and projected on the x--y plane of the simulation box. The
shortest/longest axis ratio is $s = 0.49$ ($0.48$) for the isodensity
shell (tensor ellipsoid).  The semi-major axis of the isodensity shell
$0.23 R_{\rm vir}$, consistent with JS for $2500\rho_c$ isodensity.
\label{fig:JSshells}}
\end{figure}

\citet{jing_suto02} (JS) studied 12 high resolution clusters with $N \sim 10^6$
particles and five cosmological simulations with $N = 512^3$ particles in a
$100 \hMpc$ box with both an SCDM and {\LCDM} cosmology.  The simulations were
performed with a P$^3$M code with fixed timestepping and a spatial resolution
of $10 - 20 \hkpc$.  They used a FOF halo finder and analysed the shapes of the
high resolution clusters in isodensity shells as a function of radius, finding
that the halos are more spherical at larger radii.  After determining the
relationship of shape with radius they developed a generalised ellipsoidal NFW
density profile (Figure \ref{fig:JSshells} shows that our method gives similar
results).  They applied this generalised fitting routine to the cosmological
simulations and determined generalised NFW parameters and shapes statistics.
The shapes were determined using an isodensity shell at an over density of
$2500\rho_{\rm c}$ (where $\rho_{\rm c}$ is the critical density) which
corresponds roughly to $R = 0.3R_{\rm vir}$.   The mass range analysed only
covered one order of magnitude in mass ($2.1 \times 10^{13} \hMsun \leq M_{\rm
vir} \lesssim 1 \times 10^{14} \hMsun$).  They found a result very similar to
ours, $\langle s \rangle(M_{\rm vir}) = 0.54(M_{\rm
vir}/M_*)^{-0.07[\Omega_m(z)]^{0.7}}$, and dispersion which is well fit by a
Gaussian distribution with $\sigma_s(M_{\rm vir}) = 0.113(M_{\rm
vir}/M_*)^{-0.07[\Omega_m(z)]^{0.7}}$.  We do not find any evidence for a
steepening of the exponent with redshift as they do.

The disagreement between our findings and those of JS regarding the scaling of
$\langle s \rangle$ with mass is due to the procedures used.  In the JS
analysis they determine the shape of an isodensity shell at $2500\rho_{\rm c}
\pm 3\%$, completely ignoring the interior of the shell.  JS analysed halos
with masses greater than $6.2 \times 10^{12} \hMsun$ which tend to be
dynamically young and often have double cores.  This can affect the shape a
lot, but their analysis would not pick this up, due to the neglect of the
central region.  In our iterative inertia tensor analysis we include the
centres.  In order to confirm that the difference is truly due to the shell
versus the solid ellipsoid we analysed halos from the L$200_{0.9}$ simulation
in the mass range $1-4 \times 10^{14} \hMsun$ using the technique presented in
JS.  We examined isodensity shells at $2500\rho_{\rm c}$ with a thickness of
$\pm 30\%$, instead of the $\pm 3\%$ used by JS.  We needed to examine thicker
shells in order to obtain enough particles to do the analysis because the
L$200_{0.9}$ has less mass resolution than the simulations studied by JS.  We
found that the inertia tensor method gives $\langle s \rangle_{tensor} = 0.485
\pm 0.008$ and $\sigma_s = 0.091 \pm 0.006$ and the JS method gives $\langle s
\rangle_{JS} = 0.515 \pm 0.008$ with the same scatter for the same halos.   The
difference is due to the fact that $s_{JS}$ is systematically larger at low
$s_{tensor}$.  This pattern is born out by a quantitative analysis. When we
split the sample at $s_{tensor} = 0.45$ (roughly where agreement begins), we
get that above $s_{tensor} = 0.45$ the samples agree quite well with $\langle s
\rangle_{tensor} = 0.571 \pm 0.007$ vs $\langle s \rangle_{JS} = 0.576 \pm
0.011$.  Whereas below $s_{tensor} = 0.45$ we get $\langle s \rangle_{tensor} =
0.424 \pm 0.007$ vs $\langle s \rangle_{JS} = 0.472 \pm 0.008$.  The difference
at the low end is due to the missing of the dynamically active cores by JS.  If
JS had extend their analysis to lower mass halos were multiple cores are not as
common their determination of $\langle s \rangle$ would converge with ours.
Because X-ray observers normally do not choose to only analyse the outer shells
of clusters due to the fact that the X-ray observations get noisier with the
distance from the centre and because optical observers may not see the multiple
cores when analysing cluster member velocities, we prefer the method which
includes the effect of the multiple peaks.  In Paper II we show that using our
method with some additional assumptions one can account for the observed X-ray
ellipticity measurements.

\citet{bullock02} analysed the shapes of halos in a {\LCDM} simulation with
$\se = 1.0$ at three different redshifts ($z = 0.0,1.0,3.0$).  The simulations
were performed using the ART code in a $60 \hMpc$ box with $256^3$ particles
and spatial resolution of $1.8 \hkpc$.  The analysis of shape was done using
the weighted inertia tensor in a spherical window with $R = R_{\rm vir}$.  The
axial ratios were determined iteratively until convergence was obtained using a
similar criterion as we have used, but the window remained spherical.  The use
of the weighted inertia tensor and iterative axial ratio determination seemed
to almost eliminate the effect of using a spherical window (discussed below).
\citet{bullock02} found that $\langle s \rangle(M_{\rm vir}) \simeq 0.7 (M_{\rm
vir}/10^{12} {\hMsun})^{-0.05}(1+z)^{-0.2}$ was a good fit to the simulation.
The empirical scaling of \citet{bullock02} is similar to what we find, but the
powerlaw is steeper.  This can be attributed to the lower resolution and
possibly the use of a spherical window.  Bullock's higher normalisation is due
to the higher $\se$.

\citet{kasun_evrard04} determined the shapes of cluster halos ($M_{\rm 200} > 3
\times 10^{14} \hMsun$) in the Hubble Volume simulation.  They calculated the
axial ratios using the unweighted inertia tensor in a spherical window at
$R_{200}$, the radius of the sphere within which the mean density is
$200\rho_c(z)$, with $\rho_c(z)$ being the critical density at redshift $z$.
They determined a relationship of $\langle s \rangle(M_{\rm vir}) =
0.631[1-0.023 \ln(M_{\rm vir}/10^{15} \hMsun)](1 + z)^{-0.086}$.  We compare
our analysis with theirs by performing the same spherical analysis at $R =
R_{\rm vir}$, which is slightly larger than $R_{200}$.  We find that in our
largest box simulation (L$200_{0.9}$) where we have good statistics on cluster
mass halos we find good agreement.  In examining the other two simulations for
lower mass halos we are unable to recover the extrapolation of the
\citet{kasun_evrard04} relationship.  In fact, we see a transition from the
\citet{kasun_evrard04} relationship to the relationship of
\citet{bailin_steinmetz04} (discussed below).  We also find that the mean shape
relationship has almost no dependence on radius when using a spherical window
function.

\citet{bailin_steinmetz04} analyse the shapes of halos at different radii in
spherical shells in the mass range of $10^{11} \hMsun < M \lesssim 5 \times
10^{13}$.  After determining the axial ratios they then apply an empirical
correction of $s_{true} = s_{measure}^{\sqrt{3}}$ to correct for the use of a
spherical window.  They find that all halos have an axial ratio of $\langle s
\rangle \sim 0.63$ at $R=0.4R_{\rm vir}$ with the scaling applied, which
implies that they measure $\langle s \rangle \sim 0.766$ in their spherical
window.  This result is in very good agreement with our spherical analysis
(green and red data points in Figure \ref{fig:compared}).  However, we do not
find that halos of different masses have the same mean axial ratio. There seems
to be some evidence that the $\langle s \rangle(M_{\rm vir})$ relationship
flattens out below $M_{*}$, but it is definitely not constant with radius.
Simulations with even higher mass resolution are needed to investigate for the
possibility of flattening below $M_{*}$.  For an extra check we also analysed
the halos in a spherical shell between $0.25R_{\rm vir}$ and $0.4R_{\rm vir}$
and measure a roughly flat value for all halos of $s = 0.77$.  The disagreement
about the $\langle s \rangle(M)$ relationship most likely lies in the
determination of the empirical spherical window correction.  The correction was
determined using Monte Carlo halos with no substructure, but we find that
substructure plays a role in the determined shape of the halo.

\section{Comparison with Observations}
\label{sec:obs}

\begin{figure}
\centerline{\epsfxsize=3.8in \epsffile{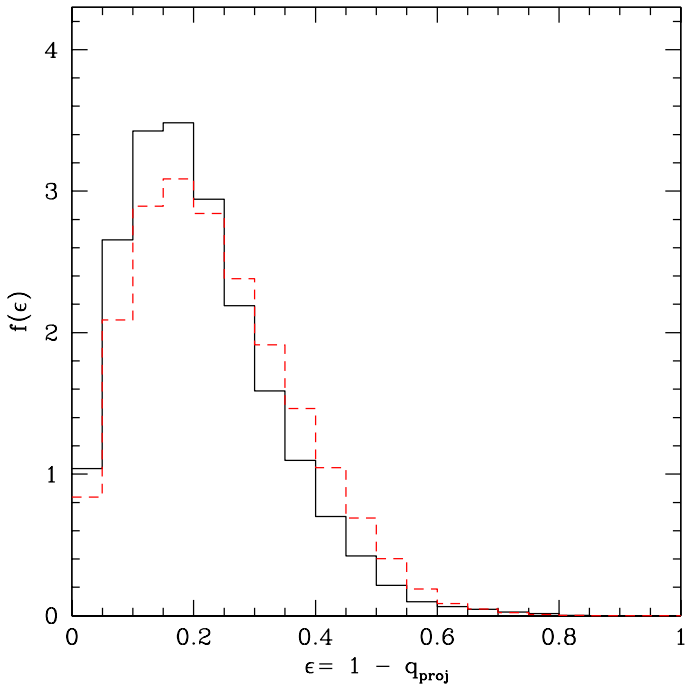}}
\caption{Projected ellipticity of galaxy mass halos at $z = 0.33$ for
$\se = 0.9$ (black solid) and $\se = 0.75$ (red dashed).
\label{fig:proj}}
\end{figure}

Since all of the differences between the shape statistics extracted from pure
collisionless simulations by various authors can be reconciled by considering
the different methods used to determine shapes, a comparison between
observations and simulations is in order.  Much of the attention halo shapes
have received lately is due to the recent estimates of the shape of the Milky
Way's host halo.  Most estimates find the Milky Way's host halo to have an
oblate shape with $s \geq 0.8$.  This is in contrast with $s \approx 0.6 \pm
0.1$ for $10^{12} \hMsun$ halos found in pure collisionless simulations, though
there is some evidence that the halos become more spherical when baryonic
cooling is included \citep{kazantzidis_etal04b,bailin_etal05} and that some
become oblate.   The presence of gas cooling will invariably make the halos
more spherical but the extent of the effect is not yet fully understood.
Recently, there have however been studies of the M giants in the leading edge
of the Sagittarius dwarf stream \citep{helmi04b,law_etal05}, which concluded
that the best fit shape of the host halo is a prolate ellipsoid with $s = 0.6$.

Another way of measuring the shape of DM halos is through weak lensing.
\citet{hoekstra_etal04} and \citet{mandelbaum_etal05} performed studies of
galaxy-galaxy weak lensing using the Red-Sequence Cluster Survey and the Sloan
Digital Sky Survey respectively.  \citet{hoekstra_etal04} determine the average
shapes of halos by measuring the orientation of the galaxies, then stacking the
galaxy images with the orientations aligned, and finally measuring the shear
field around the stacked image.  This measurement of the shear provides a rough
estimate of the dark matter halo shapes at $z \approx 0.33$.  They found an
average projected ellipticity of $\langle \epsilon \rangle \equiv \langle 1 -
q_{2D} \rangle = 0.20^{+0.04}_{-0.05}$, corresponding to $s =
0.66^{+0.07}_{-0.06}$, for halos with an average mass of $8 \times 10^{11}
\hMsun$.   The \citet{hoekstra_etal04} determination was hindered by the fact
that the galaxies were stacked together regardless of morphological type, and
one would reasonably assume that morphological type is related to merger
history and possibly orientation with the host halo which will in turn affect
the measured halo shapes as we have shown.  

In the \citet{mandelbaum_etal05} analysis of SDSS galaxies colour was used as a
proxy for morphological type.  \citet{mandelbaum_etal05} studied 2 million
lens galaxies with $r > 19$ and 31 million source galaxies dividing the lenses
into bins of colour and luminosity.  They find a suggestion that spiral (blue
galaxy) light ellipticities may be anti-aligned with the halo ellipticities at
the $2\sigma$ level and a suggestion that elliptical (red galaxy) host halo
ellipticities are luminosity dependent.  Since we and others have shown that
halo angular momentum is highly aligned with the smallest axis, this finding would
suggest that the angular momentum of spiral galaxies is not aligned with the
angular momentum of the their host halos.  Recent theoretical work by
\citet{bailin_etal05} seems to support this idea as well, although they only
study eight spiral galaxy simulations.  If this were the case one would assume
that in the \citet{hoekstra_etal04} analysis the measured signal would be
diminished by this.  It would not be completely nullified because the
\citet{hoekstra_etal04} sample is dominated by elliptical galaxies.  Indeed,
\citet{mandelbaum_etal05} show by combining the appropriate luminosity and
colour bins that their findings are in agreement with those of
\citet{hoekstra_etal04}. In Figure \ref{fig:proj} we have plotted the
distribution of projected axial ratios for the L$80_{0.9a}$ and L$80_{0.75}$
simulations for 1000 random lines of sight through the box for galaxy mass
halos.  In projection the differences between the simulations become small
hindering any sort of determination of $\se$ via lensing studies.  The peaks of
the distributions are both broadly in agreement with the findings of
\citet{hoekstra_etal04} and \citet{mandelbaum_etal05}.  Further study of the
galaxy / host halo alignment relationship and galaxy morphology / halo merger
history relationship is in order to better understand and predict the
galaxy-galaxy lensing measurements.

\section{Summary and Discussion}
\label{sec:conc}

First we investigated the dependence of halo shapes on their masses.  Our halo
sample, based on six different simulations, covers three orders of magnitude in
mass from galaxy to cluster scales.  Furthermore we have examined halo shapes
as a function of redshift and $\se$.  Our analysis of shape is based on the
halo volume enclosed by $R\approx0.3R_{\rm vir}$. This particular radius is
chosen for several reasons; halos should be fairly well relaxed within this
radius and shapes measured from X-ray gas in cluster should extend to this
radius.  At this and other radii we find that the mean shape of dark matter
halos depends on the halos mass.  We find that the mean largest-to-smallest
axial ratio $s = c/a$ at radius $0.3R_{\rm vir}$ is well described by 
\begin{equation}
\langle s \rangle(M_{\rm vir},z) = (0.54 \pm 0.03) \left(\frac{M_{\rm
vir}}{M_{*}(z)}\right)^{-0.050 \pm 0.003} 
\label{eq:powerConc}
\end{equation}
The distribution of $s$ in each mass bin is well fit by a Gaussian with $\sigma
= 0.1$.  The relation found here is steeper than that of \citet{jing_suto02} at
$z=0$, thus predicting less spherical cluster mass halos and more spherical
galaxy mass halos. 

In order to reconcile our results with sometimes strongly deviating findings by
other authors we have applied their particular methods. We find that the
disagreements between different studies are mostly due to differences between
methods of measuring axial ratios.  However, there remains one open question
brought up by this comparison.  There seems to be an inconsistency between the
analysis by \citet{kasun_evrard04}, if extended to galaxy mass halos, and
\citet{bailin_steinmetz04}.  Both groups analysed halo shapes within a
spherical window.  We find that the application of a spherical window leads to
a transition from a power-law like behaviour above $M_{*}$ to a mass
independent shape below $M_{*}$. This transition is found in the gap between
the mass ranges analysed by the two groups and therefore was not recognised by
either of them.  Possibly a similar transition can be found applying our method
of an iterative ellipsoidal window. But the effect seems to be smaller and
requires, if apparent, even higher mass resolution than utilised in this work. 

We find that the mean shape relation becomes shallower and more spherical at
increased radius. We also find that higher mass halos have a steeper
relationship between shape and radius than smaller mass halos.  Since cluster
sized halos are on average younger than galaxy sized halos, we are comparing
dynamically different objects.  The presence of an increased amount of
massive substructure near the centre of dynamically young objects may be the
reason for the steeper relation of shape with radius for cluster mass halos
than galaxy mass halos.

Our analysis of the halo shapes as a function of redshift leads to the
following results. Within fixed mass bins the redshift dependence of $\langle s
\rangle(M_{\rm vir})$ is well characterised by the evolution of $M_{*}$, unlike
the findings of \citet{jing_suto02} who predict a much steeper relation of
$\propto M_{\rm vir}^{-0.07}$ at high redshifts. We find that Equation
(\ref{eq:powerConc}) works well for different values of $\se$ (a variation of
$\se$ results in a variation of $M_{*}$ which appears as a normalisation
parameter in the $\langle s \rangle(M_{\rm vir})$ relation). Also worth noting
is that at higher redshift the possible broken power-law behaviour disappears,
but if it were truly due to $M_{*}$ we would expect this, because already by
$z=1$ $M_{*}$ is below our mass resolution. 
 
We find that the $\langle s \rangle(M_{\rm vir})$ at $z=0$ for galaxy mass
halos is $\sim 0.6$ with a dispersion of $0.1$.  This result is in good
agreement with only one estimate for the axial ratio of the Milky Way (MW)
halo.  \citet{helmi04b} claims that a study of the M giants in the leading edge
of the stream tidally stripped from the Sagittarius dwarf galaxy leads to a
best fit prolate halo with $s = 0.6$. After analysis of the same data,
\citet{law_etal05} confirm this finding. Other studies
\citep{ibata_etal01,majewski_etal03,martinez_delgado_etal04} which examined
different aspects of these streams concluded that the MW halo is oblate and
nearly spherical with $q \gtrsim 0.8$.  If the shape of the MW halo is truly
this spherical, it is either at least $2\sigma$ more spherical than the median,
or else baryonic cooling has had a large effect on the shape of the dark matter
halo (see e.g., \citealt{kazantzidis_etal04b}).  

Describing halo shapes by the triaxiality parameter $T$ introduced by
\cite{franx_etal91}, we find that the majority of halos are prolate with the
fraction of halos being prolate increasing for halos with $M_{\rm vir} >
M_{*}$. Since halo shapes are closely connected to their internal velocity
structure, we compute the angular momentum and the velocity anisotropy tensor
and relate them to both the orientation of the halo and the triaxiality.  In
agreement with previous studies we find that the angular momentum is highly
correlated with the smallest axis of the halo and that the principal axes of the
velocity anisotropy tensor tend to be highly aligned with the principal axes
of the halo. The strong alignment of all three axes of the two tensors is
remarkable since the velocity tensor tends to be more spherical, thus the
determination of its axes might be degenerate which would disturb the
correlation with the spatial axes. If the accretion of matter determines the
velocity tensor the tight correlation between velocity tensor and density shape
argues for a determination of the halo shape by merging processes.    

Finally we examine the evolution of shapes by following the merger trees of the
individual halos. We find that the different mass accretion histories of halos
cannot fully explain the observed dispersion about the mean $s$ within fixed
mass bins. It is likely that an analysis of the three dimensional accretion is
essential for the explanation of the dispersion at a fixed value of mass and
$a_c$.  However, halos with earlier formation times (lower $a_c$) tend to be
more spherical at $z=0$. Furthermore, there is a pattern of halos becoming
spherical at a more rapid rate for halos that formed earlier and this rate
appears to be independent of the final mass. The evolution of the shape for $a
> a_c + 0.1$ is well described by 
\begin{equation}
\langle s \rangle(a) \propto (a-a_c)^{\nu}\ , 
\end{equation}
where $\nu = 1.74 \times a_c^{-0.3}$. We detect a definite trend for the
transformation from highly aspherical to more spherical halo shapes after
$a_c$.  The change of $s$ seems to be less dependent on the total halo mass but
strongly influenced by the relative mass increase since $a_c$ which suggests
that halos are becoming more spherical with time due to a change in the
accretion pattern after $a_c$ from a directional to a more spherical mode. 

\section*{Acknowledgements}
We thank Anatoly Klypin for running some of the simulations which are used in
this work and for help with the others.  We also thank Volker Springel and Eric
Hayashi for helpful private communications.  The L$120_{0.9r}$ and L$80_{0.9b}$
were run on the Columbia machine at NASA Ames.  The L$80_{0.9a}$, L$80_{0.75}$,
and L$120_{0.9}$ were run on Seaborg at NERSC.  BA and JRP were supported by a
NASA grant (NAG5-12326) and a National Science Foundation (NSF) grant
(AST-0205944). AVK was supported by the NSF under grants AST-0206216 and
AST-0239759, by NASA through grant NAG5-13274, and by the Kavli Institute for
Cosmological Physics at the University of Chicago.  RHW was supported by NASA
through Hubble Fellowship HF-01168.01-A awarded by Space Telescope Science
Institute.  JSB was supported by NSF grant AST-0507816 and by startup funds at
UC Irvine.  This research has made use of NASA's Astrophysics Data System
Bibliographic Services.

\appendix

\section{Resolution Tests}
\label{app:resolution}

\begin{figure}
\centerline{\epsfxsize=3.8in \epsffile{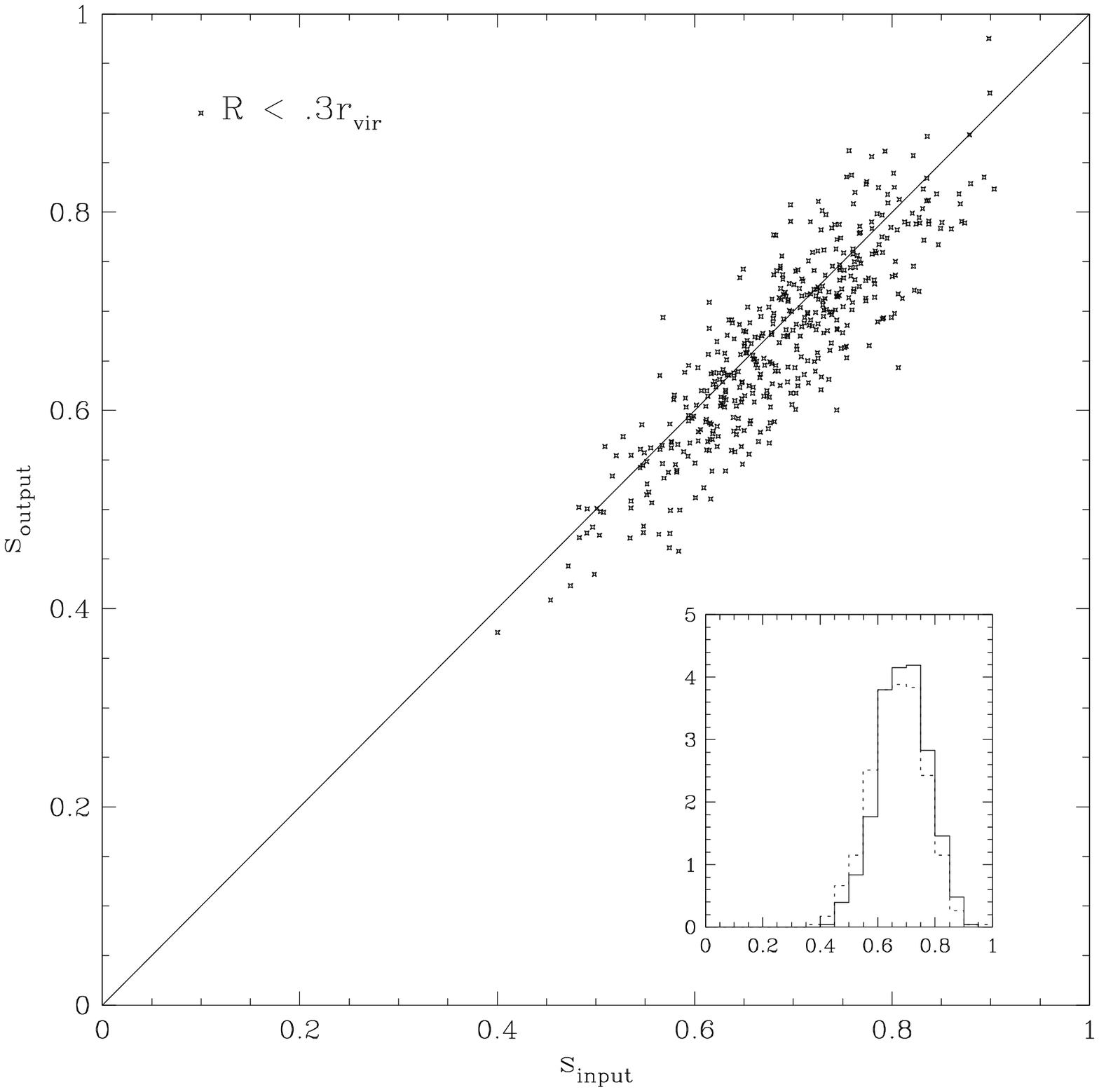}}
\caption{Results of applying our shape determination procedure at $0.3R_{\rm
vir}$ to 450 Monte Carlo halos produced with determined axial ratios.   We
found that the error in the recovered value of s could be as large as $\sim
0.1$.}
\label{fig:modelComp}
\end{figure}

\begin{figure}
\centerline{\epsfxsize=3.8in \epsffile{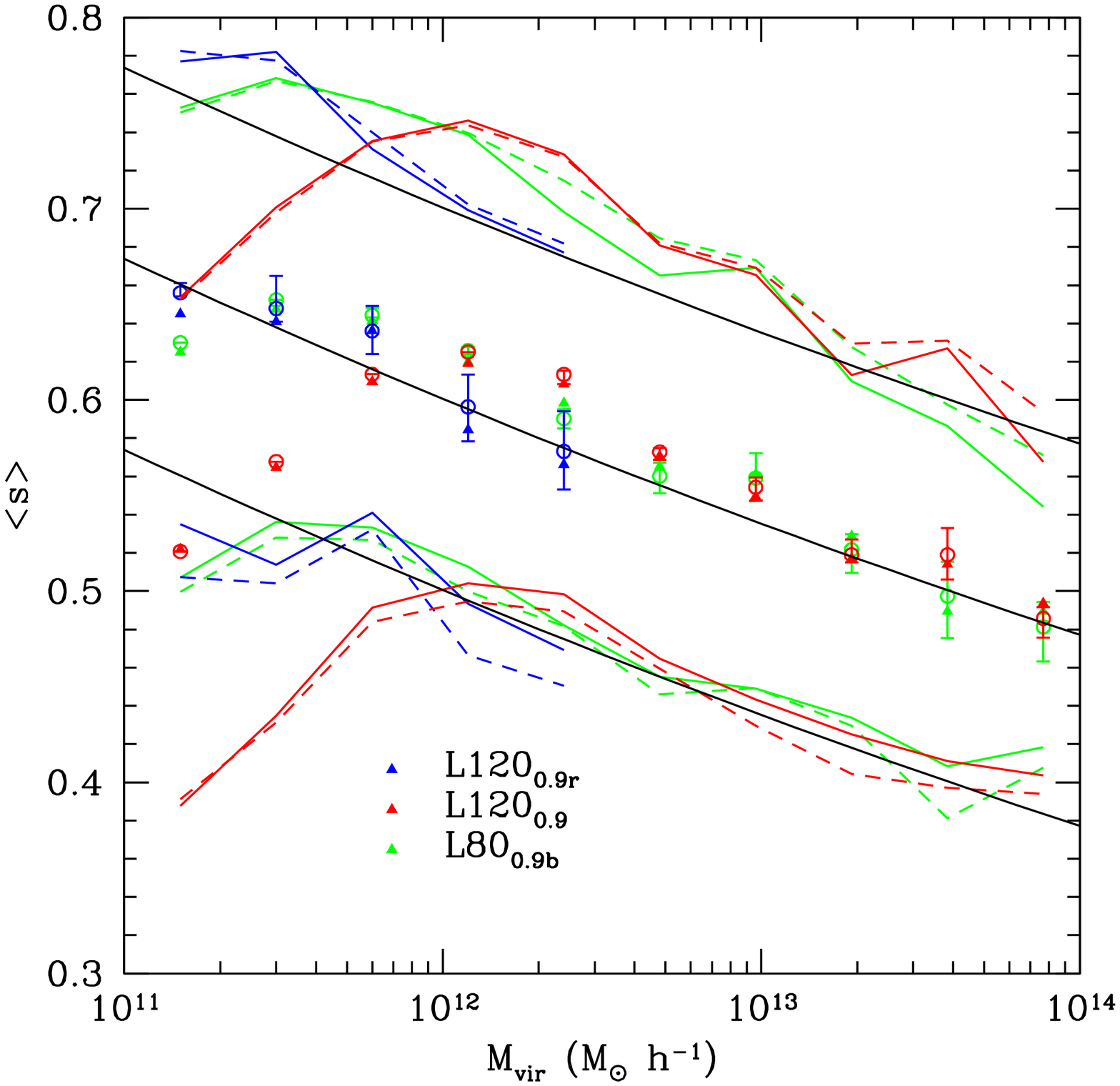}}
\caption{$\langle s \rangle$ versus mass.  This plot is a replica of Figure
\ref{fig:means} except we show mass bins below the determined resolution limit.
\label{fig:resolution}}
\end{figure}

\begin{figure}
\centerline{\epsfxsize=3.8in \epsffile{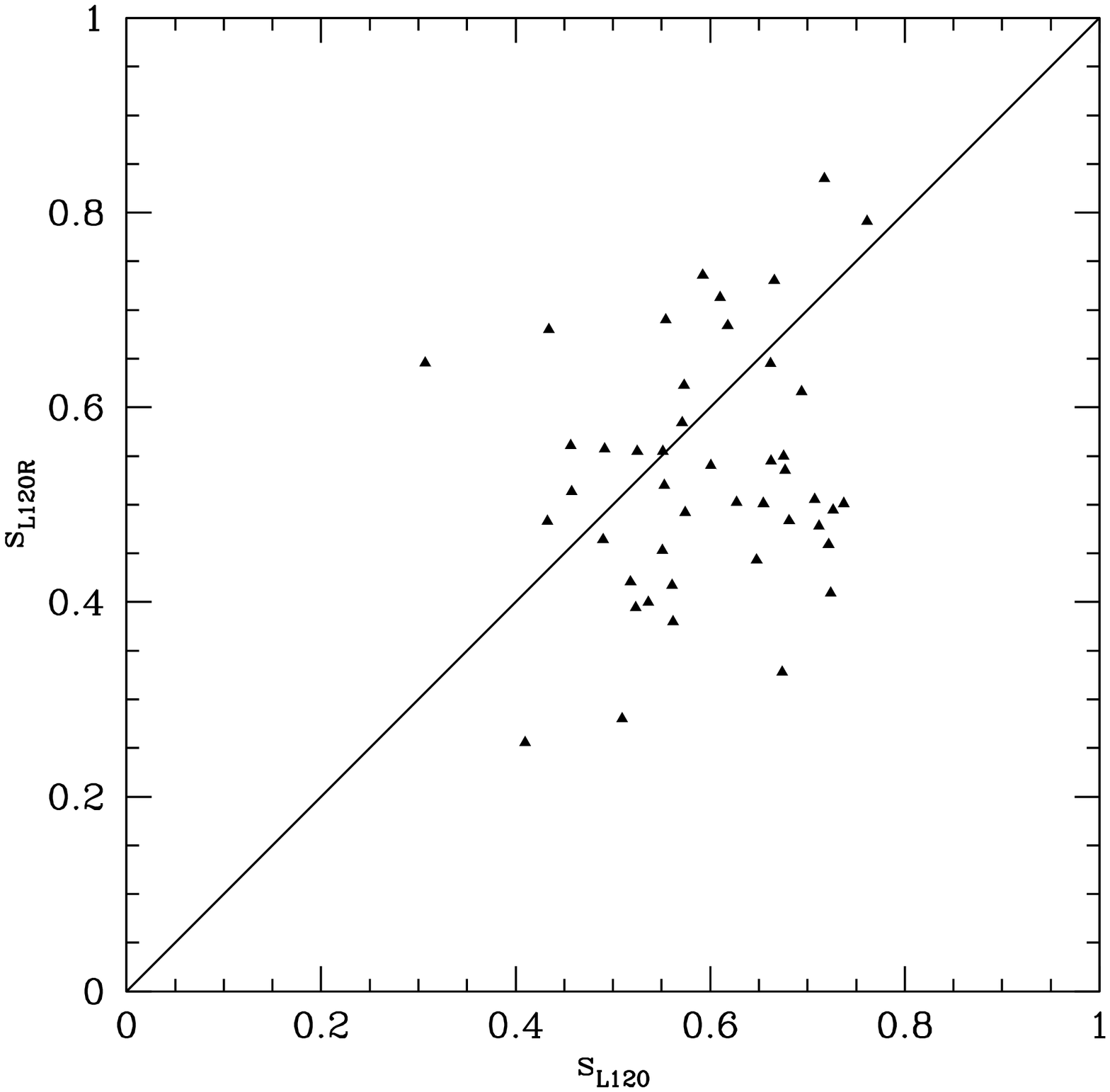}}
\caption{ A direct comparison of the axial ratios of the most massive halos in
the L$102_{0.9r}$ simulation to the corresponding halos in the L$120_{0.9}$
simulation.
\label{fig:comp120}}
\end{figure}

A potential source of systematic error in the determination of halo shapes is
the limited number of particles involved for low mass halos.  In Figure
\ref{fig:modelComp} we show the result of a Monte Carlo (MC) test.  Our dark
matter halos are adequately described by an elliptical NFW density profile
 independently of cosmological model.  Therefore, in the
figure we show the result of applying the reduced tensor method of Equation
(\ref{eq:inertia2}) to MC halos built to have an ellipsoidal NFW profile.  The
axial ratios of the MC halos are drawn from Gaussian distributions of mean
$\langle s \rangle =0.7$ and $\langle q \rangle =0.85$, and dispersion
$\sigma=0.1$.  Approximately $450$ halos were generated, each having $\sim
1000$ particles, in order to have a catalogue comparable to the sample of halos
in the L$80_{0.9}$ box with mass in the range $M_{\rm vir} = 10^{11.3} -
10^{11.7} \hMsun$.  The scatter plot shows that individual values of $s$
determined at $0.3R_{\rm vir}$ which contains roughly 300 of the 1000 particles
in the halo.  The recovered shape can be in error by up to $\sim 0.1$. However,
the scatter and mean of the distribution are very well determined by the
inertia tensor method. The inset shows the histogram for the input values of
the MC halos (solid line), and the histogram for the output values (i.e.
determined by the tensor method; dotted line). Therefore, we conclude that the
tensor method underestimates $s$ by only $0.03$ for halos of this mass. For
halos of mass $M_{\rm vir} \sim 10^{11.9} \hMsun$, the error falls to $0.01$.

The one problem with the above test is that we are attempting to recover shapes
from smooth halos.  In reality halos have substructure and the substructure
must play a role in the shape of the halo.  The presence of dense lumps close
to the core will bias $\langle s \rangle$ to lower values relative to values
determined from isodensity shells.  In order to test our ability to recover the
axial ratios in a cosmological simulation we determine the mean shape of halos
down to very low masses for all of our simulations and then compare them to one
another.  The main difference between this test and the previous one is that
the halos in the cosmological simulations contain substructure, but we do not
know a priori what the distribution of shapes should be.  In Figure
\ref{fig:resolution} we have plotted the shapes of halos down to very low
particle numbers. There appear to be two resolution effects at work here.  The
first effect is an extension of the result we found in the MC test above.  At
small particle number the recovered shape becomes very aspherical and all of
the simulations turn over at the same number of particles.  For all of the
simulations the turnover is detected at $\sim 3000$ particles within $R_{\rm
vir}$ (which is $\sim 1000$ particles within $0.3R_{\rm vir}$, where we are
determining the shape).  This is consistent with our MC tests.  At higher
particle numbers there seems to be another effect driven by the particle number
and not the mass.  Halos containing $n_p < 7000$ particles for any given
simulation show a trend of becoming more spherical on average than the
simulations of higher resolution for the same mass.  In Figure
\ref{fig:comp120} we show the determined value for $s$ for the most massive
halos shared by the L$120_{0.9}$ simulation and the resimulated subregion,
L$120_{0.9r}$.  We find that halos shapes between the two simulations can
differ by as much as $0.15$ in $s$.  This is not unexpected because in the
resimulation process the identical halos are captured at slightly different
times.  The main point here is that the average value of $s$ is lower in the
higher resolution simulation for the same halos by $\sim 0.05$.

\end{document}